\documentclass[preprint2]{aastex63}

\usepackage{enumitem}
\usepackage{hyperref}
\usepackage{multirow}
\usepackage{xcolor}

\DeclareUnicodeCharacter{2212}{-}

\shorttitle{Improved Properties for 52 Planet-Hosting Cool Dwarfs}
\shortauthors{Gore, Giacalone et al.}

\submitjournal{Astrophysical Journal Supplement Series}
\accepted{February 18, 2024}

\begin{document}

\title{Metallicities and Refined Stellar Parameters for 52 Cool Dwarfs with Transiting Planets and Planet Candidates}

\correspondingauthor{Steven Giacalone}
\email{giacalone@astro.caltech.edu}
 
\author{Rebecca Gore}
\affiliation{Department of Astronomy, University of California Berkeley, Berkeley, CA 94720, USA}
 \affiliation{Bay Area Environmental Research Institute, Moffett Field, CA 94035, USA}
\affiliation{NASA Ames Research Center, Moffett Field, CA 94035, USA}

\author[0000-0002-8965-3969]{Steven Giacalone}
\altaffiliation{NSF Astronomy and Astrophysics Postdoctoral Fellow}
\affiliation{Department of Astronomy, University of California Berkeley, Berkeley, CA 94720, USA}
\affiliation{Department of Astronomy, California Institute of Technology, Pasadena, CA 91125, USA}

\author[0000-0001-8189-0233]{Courtney D. Dressing}
\affiliation{Department of Astronomy, University of California Berkeley, Berkeley, CA 94720, USA}

\author[0000-0002-1845-2617]{Emma V. Turtelboom}
\affiliation{Department of Astronomy, University of California Berkeley, Berkeley, CA 94720, USA}

\author{Ashley Schroeder}
\affiliation{Department of Astronomy, University of California Berkeley, Berkeley, CA 94720, USA}
\affiliation{AstroCamp, Idyllwild, CA 92549, USA}

\author[0000-0001-5286-639X]{Charles D. Fortenbach}
\affiliation{Department of Physics and Astronomy, San Francisco State University, San Francisco, CA 94132, USA}

\author[0000-0003-3702-0382]{Kevin K.\ Hardegree-Ullman}
\affiliation{Steward Observatory, The University of Arizona, Tucson, AZ 85721, USA}

\author[0000-0003-1848-2063]{Jon K.\ Zink}
\altaffiliation{NHFP Sagan Fellow}
\affiliation{Department of Astronomy, California Institute of Technology, Pasadena, CA 91125, USA}

\author[0000-0002-7216-2135]{Andrew W. Mayo}
\affiliation{Department of Astronomy, University of California Berkeley, Berkeley, CA 94720, USA}
\affiliation{Centre for Star and Planet Formation, Natural History Museum of Denmark \& Niels Bohr Institute, University of Copenhagen, \O ster Voldgade 5-7, DK-1350 Copenhagen K., Denmark}

\author[0000-0001-5347-7062]{Joshua E.\ Schlieder}
\affiliation{NASA Goddard Space Flight Center, 8800 Greenbelt Rd, Greenbelt, MD 20771, USA}

\author[0000-0002-8035-4778]{Jessie L.\ Christiansen}
\affiliation{NASA Exoplanet Science Institute, IPAC, MS 100-22, Caltech, 1200 E. California Blvd, Pasadena, CA 91125, USA}

\begin{abstract}
We collected near-infrared spectra of 65 cool stars with the NASA InfraRed Telescope Facility (IRTF) and analyze them to calculate accurate metallicities and stellar parameters. The sample of 55 M dwarfs and 10 K dwarfs includes 25 systems with confirmed planets and 27 systems with planet candidates identified by the \textit{K2} and \textit{TESS} missions. Three of the 25 confirmed planetary systems host multiple confirmed planets and two of the 27 planet candidate systems host multiple planet candidates. Using the new stellar parameters, we re-fit the \textit{K2} and \textit{TESS} light curves to calculate updated planet properties. In general, our updated stellar properties are more precise than those previously reported and our updated planet properties agree well with those in the literature. Lastly, we briefly examine the relationship between stellar mass, stellar metallicity, and planetary system properties for targets in our sample and for previously characterized planet-hosting low-mass stars. We provide our spectra, stellar parameters, and new planetary fits to the community, expanding the sample available with which to investigate correlations between stellar and planetary properties for low-mass stars.

\end{abstract}

\keywords{ Exoplanets (498), Exoplanet systems (484), Stellar properties (1624), M dwarf stars (982), Low mass stars (2050), Spectroscopy (1558)}

\section{Introduction} \label{sec:intro}

The {\it Kepler} and {\it K2} missions discovered thousands of transiting planets in our galaxy.\footnote{\url{https://exoplanetarchive.ipac.caltech.edu/docs/counts_detail.html}} While these missions were not primarily concerned with planets around M dwarfs \citep{borucki2010, howell2014}, they resulted in the discovery of hundreds of planets orbiting these cool stars \citep[e.g.,][]{morton2016}. These discoveries have allowed for robust estimations of the occurrence rates of planets orbiting M dwarfs, specifically as a function of planet radius ($R_{\rm p}$) and orbital period ($P_{\rm orb}$) \citep[e.g.,][]{dressing2013, dressing2015, muirhead2015, mulders2015, gaidos2016, hardegree-ullman2019, bergsten2023}. The number of known M dwarf planetary systems has increased even further with the Transiting Exoplanet Survey Satellite \citep[{\it TESS};][]{ricker2015}, which can access low-mass stars more easily than \textit{Kepler} due to its higher sensitivity at long wavelengths. Some yield predictions projected the discovery of $\geq 1000$ new transiting planets orbiting M dwarfs, most of which will be small ($R_{\rm p} < 4 \, R_\oplus$) \citep{ballard2019, kunimoto2022}. Others have found this prediction to be over-optimistic, and that \textit{TESS} has delivered fewer small planets around cool stars than expected, possibly due to a low planet occurrence rate around the latest M dwarfs \citep{brady2022}. In either case, the sample of cool dwarf planets found by \textit{TESS} will provide a foundation on which to more robustly explore correlations between the properties of cool stars and their planetary systems.

Using this new sample of planets, we can gain a better understanding of the relationship between planet occurrence rates and stellar metallicity. It has been shown that giant planets occur more frequently around more metal-rich FGK stars \citep[e.g.,][]{fischer2005, johnson2010, sousa2011}. This correlation is more tenuous for M dwarfs, around which short-period giant planets are relatively rare and, due to the intrinsic faintness of low-mass stars, more challenging to find \citep[e.g.,][]{johnson2009, rojas-ayala2012, terrien2012, gan2022, gan2023, hartman2023}. Several studies have searched for a link between stellar metallicity and the occurrence rate of smaller planets. Using a sample of several hundred planet-hosting stars, \citet{buchhave2014} argued that stars hosting small ($R_{\rm p} < 1.7 \, R_\oplus$) close-in planets have lower metallicities than those hosting close-in sub-Neptunes ($1.7 \, R_\oplus < R_{\rm p} < 3.9 \, R_\oplus$) and gas giants ($R_{\rm p} > 3.9 \, R_\oplus$), on average. \citet{mulders2016} further explored this relationship by examining the connection between host star metallicity and planet orbital period, finding that small planets with $P_{\rm orb} < 10$ days orbit stars with higher metallicities than small planets with  $P_{\rm orb} > 10$ days. \citet{petigura2018} analyzed the link between host star metallicity and the frequency of small planets, concluding that they are equally common around metal-poor and metal-rich stars (also see \citealt{buchhave2012}). These trends, or lack thereof, are most apparent for FGK stars, but have also been tentatively identified for M dwarfs \citep{rojas-ayala2012, neves2013, hobson2018, maldonado2020}.

Some studies have also found evidence of correlations between stellar metallicity and system architecture. For instance, \citet{brewer2018} showed that compact multi-planet systems are more commonly found around FGK stars that are metal poor than those that are metal rich (although we note that other studies find no such correlation; \citealt{weiss2018}). \citet{anderson2021} later fount a similar trend as \citet{brewer2018} for M dwarfs and late K dwarfs using Gaia color as a proxy for metallicity, showing that compact multi-planet systems tend to be around cool stars that are significantly bluer (more metal-poor) than stars hosting only a single planet. In an independent analysis that utilized planet-hosting M dwarfs with photometrically and spectroscopically determined metallicities, \citet{rodriguezmartinez2023} also found a tendency for multi-planet systems to orbit more metal-poor M dwarfs than single-planet systems. As more planetary systems are discovered around these cool stars, having reliable metallicities will be crucial for similar population-level studies that aim to infer how this property influences planet formation and evolution around low-mass stars.

Our ability to measure the properties of M dwarfs has improved drastically over the last decade. In preparation for the {\it K2} mission, the properties of cool stars were estimated primarily using broadband photometry and proper motions via the {\it Galaxia} galactic model \citep{sharma2011} and Padova isochrone models \citep{girardi2000, marigo2007, marigo2008}, which were compiled in the Ecliptic Plane Input Catalog \citep[EPIC;][]{EPIC2016}. While the EPIC generally distinguishes cool dwarf stars from hotter stars, the documentation cautions that the stellar radius estimates could be systematically lower than the true values due to a tendency for the Padova models to underestimate the radii of cool dwarfs \citep{boyajian2012}. This discrepancy poses a problem for studies of transiting planets around these stars, as underestimated stellar radii lead to underestimated planet radii for any transiting planets.

In the last decade, methods have been developed to circumvent this issue by using empirical relations to determine the physical properties of cool stars from parallax measurements and broadband photometry \citep[e.g., ][]{Mann2013, Mann2014, Mann2015, Mann2019, Newton2014, Newton2015}. These relations were leveraged in the curation of a cool dwarf list in the {\it TESS} Input Catalog \citep[TIC;][]{muirhead2018, TIC2018, TIC2019}, which utilized the parallax measurements of millions of nearby M dwarfs in Gaia DR2 \citep{gaia2021}. In addition, these relations made it possible to refine stellar parameters for M dwarfs by including measurements of stellar metallicity extracted from near infrared spectra \citep[e.g.,][]{Mann2013, Mann2014, Mann2015, Mann2019, Newton2014, Newton2015, terrien2015, dressing2017, martinez2017, wittenmyer2018, dressing2019}.

In this paper we measure the metallicities of M dwarfs with transiting planet candidates and confirmed transiting planets and use them to calculate refined stellar parameters. We then use these parameters to calculate updated radii and insolation fluxes for the planet candidates and planets. This work complements other surveys that have measured the metallicities of planet-hosting M dwarfs \citep[e.g.,][]{bean2006, mann2012, Mann2013, Mann2014, Mann2015, Mann2019, Newton2014, Newton2015, terrien2015, dressing2017, dressing2019, maldonado2020, rains2021}.

This paper is structured as follows. In Section \ref{sec:sample}, we describe the sample of stars characterized in this paper and the criteria used to select them. In Section \ref{sec:observations}, we discuss our observation strategy and data reduction procedure. In Section \ref{sec:analysis}, we detail the process of extracting metallicities and improved stellar parameters from our spectra. In Section \ref{sec:planets}, we calculate updated orbital {parameters} and radii for the confirmed planets and planet candidates orbiting the stars in our sample. In Section \ref{sec:discussion}, we explore the relationship between planetary system properties and stellar parameters in our sample and for previously characterized low-mass stars. Lastly, in Section \ref{sec:conclusions}, we provide concluding remarks.

\section{Target Selection} \label{sec:sample}

The goal of this paper is to calculate the metallicities and update stellar parameters for M dwarfs and late K dwarfs (M5 -- K7) with transiting planets and planet candidates. Accordingly, we selected targets to observe based on the following criteria: (1) the star must have either a transiting planet candidate or a known transiting planet, (2) the star must have an estimated effective temperature ($T_{\rm eff}$) less than 5300 K and a surface gravity ($\log g$) greater than 4.0 according to the EPIC \citep{huber2014} or version 8.2 of the TIC (\citealt{TIC2018, TIC2019}; with a preference for lower values of $T_{\rm eff}$ and higher values of $\log g$), and (3) the star must be observable from the IRTF (that is, it must have a declination between $70^\circ$ and $-59^\circ$). In addition, we prioritized brighter stars that allowed us to acquire higher signal-to-noise (S/N) spectra as well as stars that seem scientifically compelling, either because they already had a confirmed planet or had a promising planet candidate. {Lastly, we consulted the IRTF Data Archive\footnote{\url{https://irsa.ipac.caltech.edu/applications/irtf/}} to verify that each target did not already have IRTF/SpeX data from earlier large-scale cool star surveys \citep[e.g.,][]{Newton2014, terrien2015}}. We also observed a small number of cool stars that are not flagged as having transiting planet candidates for the purpose of compiling a set of ``control'' stars with which to compare the properties of planet-hosting stars. We note that stars within the control sample are statistically likely to harbor planets in non-transiting geometries or planets that were not detectable in transit given the existing \textit{TESS} and \textit{K2} data. We include these stars in our analyses, as they may be of interest to future studies.

Our complete target sample, which consists of 65 stars, is presented in Table \ref{tab: initial_params}. This table includes a number of stellar properties from various catalogs that we recalculate in Section \ref{sec: detailed_classification}. The {two} right-most {columns} of the table {indicate} whether the {objects of interest around the stars are planet candidates, confirmed planets, or false positives, as well as the dates on which these objects were classified as such}. These flags {and dates} are based on {information} in the NASA Exoplanet Archive (\dataset[10.26133/NEA1]{https://exoplanetarchive.ipac.caltech.edu/cgi-bin/TblView/nph-tblView?app=ExoTbls&config=PS}) and {ExoFOP (\dataset[10.26134/ExoFOP3]{https://exofop.ipac.caltech.edu/}). We adopted the following key for labeling planetary systems
\begin{itemize}[noitemsep]
    \item {\textbf{CP:} The star hosts one or more confirmed planets.}
    \item {\textbf{PC:} The star hosts at least one planet candidate but does not host confirmed planets.}
    \item {\textbf{FP:} The star was previously reported to host a planet candidate that was later determined to be a false positive.}
\end{itemize}
We note that TIC 163539739 (TOI-1278) is labeled as CP but actually hosts a transiting brown dwarf \citep[$M = 18.5 \pm 0.5 \, M_{\rm J}$;][]{Artigau2021}, but we choose to keep it in our analysis due to the rarity of these objects \citep[e.g.,][]{grether2006, sahlmann2011}.Planet candidates and confirmed planets orbiting stars in our sample are discussed further in Sections \ref{sec:planets} and \ref{sec:discussion}.

\begin{longrotatetable}
\movetabledown=18mm
\begin{deluxetable*}{lllcccccccccccc}\label{tab: initial_params}
\tablecaption{Initial Properties of the 65 Systems in this Work}
\tabletypesize{\scriptsize}
\setlength\extrarowheight{0pt}
\tablewidth{\textwidth}
 \tablehead{
 \colhead{ID} & \colhead{Alt. Name} & \colhead{Campaign/Sector} &  \colhead{$V$} & \colhead{$J$} & \colhead{$Ks$} & \colhead{$d$ (pc)} & \colhead{$R_\star$ ($R_\odot$)} & \colhead{$M_\star$ ($M_\odot$)} & \colhead{$\log g$ (dex)} & \colhead{$T_{\rm eff}$ (K)} & \colhead{Sp.T.} & \colhead{S/N} & \colhead{Disp.} & Date
 }
\startdata
 EPIC 211510580 &             - &                     5,18 & 16.34 & 12.48 & 11.63 &  $99.70 \pm 0.30$ & $0.36 \pm 0.19$ & $0.39 \pm 0.21$ & $4.920\pm 0.020$ & $3855 \pm 431$ &   M2.5 V & 130 & PC & - \\  
 EPIC 211586387 &             - &                  5,16,18 & 14.64 & 12.70 &  12.04 & $323.62 \pm 2.09$ &  $0.67 \pm 0.04$ & $0.74 \pm 0.05$ & $4.647 \pm 0.036$ & $4660 \pm 111$ &   K2 V & 133 & PC & - \\  
 EPIC 211663688 &             - &                       16 & 19.11 & 14.12 &  13.25 &  $99.80 \pm 0.80$ &  $0.32 \pm 0.09$ & $0.34 \pm 0.13$ & $4.942 \pm 0.102$ & $3768 \pm 221$ &   M4.5 V & 47 & PC & - \\  
 EPIC 211673349 &             - &                       16 & 14.76 & 12.65 &  11.90 & $305.81 \pm 1.87$ &   $0.53 \pm 0.07$ & $0.59 \pm 0.09$ & $4.754 \pm 0.064$ & $4245 \pm 83$ &    K4 V & 159 & PC & - \\  
 EPIC 211731298 &             - &                     5,18 & 13.20 & 11.73 &  11.23 & $285.71 \pm 1.63$ &    $0.78 \pm 0.06$ & $0.86 \pm 0.05$ & $4.571 \pm 0.020$ & $5244 \pm 83$ &    K2 V & 169 & PC & - \\  
 EPIC 211863149 &             - &                       16 & 15.89 & 12.18 &  11.27 & $117.23 \pm 0.41$ &  $0.32 \pm 0.07$ & $0.34 \pm 0.09$ & $4.956 \pm 0.078$ & $3730 \pm 175$ &    M2.5 V  & 154 & PC & - \\  
 EPIC 211925595 &             - &                     5,16 & 14.58 & 13.02 &  12.43 & $478.47 \pm 4.58$ &  $0.72 \pm 0.05$ & $0.79 \pm 0.05$ & $4.612 \pm 0.030$ & $4874 \pm 77$ &        K2 V & 133 & PC & - \\  
 EPIC 212048748 &        K2-313 &                       16 & 13.82 & 10.06 & 9.19 &  $27.83 \pm 0.02$ &  $ 0.31 \pm 0.04$ & $0.33 \pm 0.05$ & $4.964 \pm 0.049$ & $3713 \pm 129$ &    M2.5 V  & 211 & CP & 2020-03 \\  
 EPIC 212088059 &        K2-345 &                  5,16,18 & 15.72 & 12.34 &  11.46 & $161.55 \pm 0.52$ &  $0.40 \pm 0.08$ & $0.45 \pm 0.10$ & $4.877 \pm 0.072$ & $3852 \pm 136$ &     M1 V & 159 & CP & 2021-11 \\  
 EPIC 248433930 &             - &                       14 & 14.45 & 11.54 &  10.72 & $115.87 \pm 0.27$ &   $0.38 \pm 0.08$ & $0.43 \pm 0.11$ & $4.903 \pm 0.080$ & $3829 \pm 122$ &        M1 V & 109 & -  & - \\  
 EPIC 248440276 &             - &                       14 & 14.60 & 10.93 &  10.08 &  $57.08 \pm 0.07$ &   $0.24 \pm 0.04$ & $0.23 \pm 0.06$ & $5.042 \pm 0.050$ &  $3525 \pm 69$ &      M2.5 V & 184 & -  & - \\  
 EPIC 248527514 &        K2-401 &                       14 &  14.01 & 11.87 &  11.11 & $201.61 \pm 0.81$ &  $0.40 \pm 0.13$ & $0.45 \pm 0.14$ & $4.878 \pm 0.128$ &  $4200 \pm 149$ &    K5 V & 165 &  CP  & 2022-06 \\  
 EPIC 248731669 &             - &                       14 & 16.57 & 13.38 &  12.52 & $261.78 \pm 3.43$ &  $0.31 \pm 0.08$ & $0.35 \pm 0.10$ & $4.966 \pm 0.088$ &  $3750 \pm 175$ &        M0.5 V & 102 & -  & - \\  
 EPIC 248775938 &             - &                       14 & 15.88 & 13.09 &  12.20 & $311.53 \pm 3.88$ &  $0.46 \pm 0.05$ & $0.53 \pm 0.06$ & $4.817 \pm 0.048$ & $4018 \pm 96$ &        M0.5 V & 133 & PC  & - \\  
 EPIC 248856002 &             - &                       14 & 16.00 & 13.07 &  12.15 & $189.04 \pm 1.43$ &  $0.28 \pm 0.08$ & $0.29 \pm 0.10$ & $5.014 \pm 0.096$ & $3663 \pm 128$ &        M2 V & 140 & -  & - \\  
 EPIC 248861279 &         K2-404 &                       14 & 14.62 & 11.58 &  10.75 & $121.65 \pm 0.30$ &  $0.35 \pm 0.05$ & $0.39 \pm 0.06$ & $4.928 \pm 0.048$ &  $3775 \pm 60$ &        M1 V & 161 & CP  & 2022-06 \\  
  TIC 2760710 &          TOI-227 &                     2,29 & 18.44 & 12.13 &  11.25 &  $52.63 \pm 0.11$ &   $0.24 \pm 0.01$ &  $0.21 \pm 0.02$ & $4.998 \pm 0.014$ & $3176 \pm 157$ &  M4.5 V & 116 & PC  & - \\  
TIC 415969908 &          TOI-233 &                  2,29,43 & 13.34 & 9.94 &   9.09 &  $33.73 \pm 0.02$ &    $0.38 \pm 0.01$ &  $0.37 \pm 0.02$ & $4.845 \pm 0.002$ & $3568 \pm 157$ &  M2.5 V & 165 & 2$\times$PC  & - \\  
 TIC 12423815 &          TOI-234 &                     2,29 & 16.52 & 13.22 &  12.33 & $253.16 \pm 2.56$ &   $0.55 \pm 0.02$ &  $0.54 \pm 0.02$ & $4.695 \pm 0.011$ & $3628 \pm 157$ &     M0.5 V & 98 & PC  & - \\  
TIC 118327550 &          TOI-244 &                     2,29 &  12.86 & 8.83 &   7.97 &  $22.08 \pm 0.01$ &  $0.41 \pm 0.01$ &  $0.40 \pm 0.02$ & $4.820 \pm 0.004$ & $3407 \pm 157$ &     M2.5 V & 70 & CP  & 2023-07 \\  
TIC 120916706 &          TOI-263 &                   3,4,30 &  18.97 & 14.08 &  13.25 & $286.53 \pm 5.75$ &  $0.44 \pm 0.02$ & $0.44 \pm 0.02$ & $4.791 \pm 0.010$ & $3423 \pm 157$ &        M3 V & 58 & FP  & 2021-03 \\  
 TIC 70899085 &          TOI-442 &                     5,32 &  12.49 & 9.49 &   8.63 &  $52.33 \pm 0.05$ &  $0.61 \pm 0.02$ & $0.59 \pm 0.02$ & $4.645 \pm 0.011$ & $3779 \pm 157$ &       M1 V & 207 & CP  & 2020-12 \\  
TIC 365639282 &          TOI-482 &                     6,32 &  14.94 & 11.74 &  10.95 & $143.06 \pm 4.09$ &  $0.56 \pm 0.02$ & $0.56 \pm 0.02$ & $4.684 \pm 0.011$ & $3692 \pm 157$ &      M2.5 V & 126 & FP  & 2019-11 \\  
TIC 452866790 &          TOI-488 &                  7,34,61 & 13.74 & 9.63 &   8.83 &  $27.31 \pm 0.01$ &  $0.35 \pm 0.01$ & $0.34 \pm 0.02$ & $4.870 \pm 0.001$ & $3329 \pm 157$ &       M3.5 V & 201 & 2$\times$CP  & 2020-10 \\  
 TIC 19025965 &          TOI-493 &           7,34,44--46,61 &   12.55 & 10.36 &   9.63 & $107.41 \pm 0.23$ &  $0.81 \pm 0.07$ & $0.65 \pm 0.03$ & $4.431 \pm 0.098$ & $4139 \pm 106$ &  M0 V  & 216 & PC  & - \\  
TIC 348538431 &          TOI-507 &                  7,34,61 &  16.30 & 11.65 &  10.84 & $109.89 \pm 0.24$ &  $0.51 \pm 0.02$ & $0.50 \pm 0.02$ & $4.733 \pm 0.009$ & $3330 \pm 157$ &  M4 V & 159 & PC  & - \\  
TIC 318937509 &          TOI-516 &                        7 &  17.20 & 12.36 &  11.49 & $152.67 \pm 0.93$ &  $0.51 \pm 0.02$ & $0.50 \pm 0.02$ & $4.732 \pm 0.009$ & $3225 \pm 157$ &  M0.5 V & 101 & FP  & 2020-11 \\  
TIC 218795833 &          TOI-519 &                   7,8,34 &  17.35 & 12.85 &  11.95 & $115.21 \pm 0.53$ &  $0.36 \pm 0.01$ & $0.34 \pm 0.02$ & $4.869 \pm 0.001$ & $3324 \pm 157$ &     M2.5 V & 102 & CP  & 2023-08 \\  
 TIC 27649847 &          TOI-521 &           7,34,44--46,61 &   14.69 & 10.92 &  10.09 &  $61.01 \pm 0.07$ &  $0.42 \pm 0.01$ & $0.41 \pm 0.02$ & $4.808 \pm 0.005$ & $3465 \pm 157$ &   M2 V & 270 & PC  & - \\  
 TIC 93125144 &          TOI-523 &                  6,32,33 &   10.71 & 8.74 &   8.12 &  $69.93 \pm 0.83$ &  $0.89 \pm 0.05$ & $0.79 \pm 0.09$ & $4.434 \pm 0.065$ & $4852 \pm 121$ &                                K2 V & 219 & FP  & 2021-03 \\  
TIC 200593988 &          TOI-526 &                        6 &   14.31 & 10.92 &  10.11 &  $70.87 \pm 0.10$ &  $0.47 \pm 0.01$ & $0.46 \pm 0.02$ & $4.767 \pm 0.007$ & $3601 \pm 157$ &    M1.5 V & 447 & FP  & 2022-10 \\  
TIC 387690507 &          TOI-530 &               6,33,44,45 &   15.40 & 12.11 &  11.24 & $147.71 \pm 0.65$ &  $0.54 \pm 0.02$ & $0.54 \pm 0.02$ & $4.698 \pm 0.010$ & $3566 \pm 157$ &     M1 V & 114 & CP  & 2022-03 \\  
TIC 144700903 &          TOI-532 &                        6 &   14.41 & 11.47 &  10.59 & $135.32 \pm 0.37$ &   $0.62 \pm 0.02$ & $0.61 \pm 0.02$ & $4.631 \pm 0.011$ & $3815 \pm 157$ &      M0V  & 126 &  CP  & 2021-10 \\  
TIC 237751146 &           TOI-538 &                    6,33 &  16.54 & 12.62 &  11.70 & $134.23 \pm 0.90$ &  $0.43 \pm 0.01$ & $0.43 \pm 0.02$ & $4.796 \pm 0.006$ & $3352 \pm 157$ &    M3 V    & 99 &  FP  & 2021-02 \\  
TIC  50618703 &           TOI-544 &                    6,32 &    10.78 & 8.59 &  7.80 &  $40.92 \pm 0.03$ &  $0.66 \pm 0.06$ & $0.66 \pm 0.08$ & $4.614 \pm 0.110$ & $4220 \pm 123$ &    K7 V    & 244 & CP  & 2022-02 \\  
TIC 192826603 &           TOI-551 &                5,6,32,33 & 15.67 & 12.70 &  11.94 & $217.86 \pm 0.95$ &  $0.57 \pm 0.02$ & $0.56 \pm 0.02$ & $4.679 \pm 0.010$ &  $3902 \pm 157$ & M0.5 V      & 109 &  CP  & 2018-04 \\  
TIC 170849515 &           TOI-555 &                  5,31,32 &  16.68 & 13.69 &  12.81 & $395.26 \pm 4.69$ &  $0.63 \pm 0.02$ & $0.61 \pm 0.02$ & $4.623 \pm 0.013$ & $3929 \pm 157$ &    M1.5 V  & 78 & CP  & 2022-03 \\  
 TIC 55488511 &           TOI-557 &                     5,31 &   13.34 & 10.41 &   9.61 &  $75.47 \pm 0.11$ &  $0.57 \pm 0.02$ & $0.57 \pm 0.02$ & $4.673 \pm 0.010$ & $3841 \pm 157$ &   K7 V  & 178 &  PC   & - \\  
TIC 101011575 &           TOI-560 &                  8,34,61 &    9.67 & 7.65  &  6.95 &   $31.59 \pm 0.02$ &  $0.63 \pm 0.05$ & $0.75 \pm 0.09$ & $4.714 \pm 0.010$ & $4695 \pm 138$ &  K2 V   &  299 & 2$\times$CP  & 2022-08 \\  
TIC 413248763 &           TOI-562 &                  8,35,62 &   10.91 & 7.34 &  6.47 & $9.436 \pm 0.002$ &  $0.36 \pm 0.01$ & $0.35 \pm 0.02$ & $4.864 \pm 0.001$ & $3490 \pm 157$ &  M2 V & 323 & 3$\times$CP   & 2019-08 \\  
  TIC 1133072 &           TOI-566 &            8,34,35,61,62 &   15.34 & 11.17 &  10.28 &  $84.18 \pm 0.14$ &  $0.50 \pm 0.01$ & $0.49 \pm 0.02$ & $4.741 \pm 0.008$ &  $3471 \pm 157$ & M2.5 V  & 110 & FP  & 2019-04 \\  
TIC 296780789 &           TOI-573 &                  8,35,62 &   15.02 & 11.11 &  10.23 &  $88.73 \pm 0.16$ &   $0.34 \pm 0.06$ & $0.34 \pm 0.07$ & $4.911 \pm 0.395$ &    -   &          M2 V & 70 & PC & - \\  
TIC 296739893 &           TOI-620 &                  8,35,62 &   12.26 & 8.84 &   7.95 &  $33.01 \pm 0.03$ &   $0.55 \pm 0.02$ & $0.54 \pm 0.02$ & $4.696 \pm 0.010$ & $3586 \pm 157$ &    M2.5 V & 113 & CP  & 2022-06 \\  
 TIC 32497972 &           TOI-876 &                5,6,32,33 &   13.18 & 10.33 &   9.51 &  $78.06 \pm 0.06$ &  $0.60 \pm 0.02$ & $0.59 \pm 0.02$ & $4.648 \pm 0.011$ & $3882 \pm 157$ &    M0.5 V & 157 & PC  & - \\  
 TIC 29960110 &          TOI-1201 &                     4,31 &   12.26 & 9.53 &   8.65 &  $37.63 \pm 0.02$ &  $0.48 \pm 0.01$ & $0.48 \pm 0.02$ & $4.753 \pm 0.008$ & $3506 \pm 157$ &    M2 V & 246 & CP  & 2021-12 \\  
TIC 163539739 &          TOI-1278 &                 15,55,56 &   13.40 & 10.60 &   9.73 &  $75.47 \pm 0.06$ &  $0.55 \pm 0.02$ & $0.55 \pm 0.02$ & $4.694 \pm 0.009$ & $3841 \pm 157$ &  M0.5V    & 164 & CP  & 2021-10 \\  
 TIC 13684720 &          TOI-1433 &              14,15,41,55 &   15.60 & 10.81 &   9.95 &  $68.07 \pm 0.05$ &  $0.48 \pm 0.01$ & $0.47 \pm 0.02$ & $4.758 \pm 0.007$ & $3275 \pm 157$ &     M3V & 159 & PC  & - \\  
TIC 343628284 &          TOI-1448 &                15--17,57 &   15.68 & 11.60 &  10.77 &  $73.64 \pm 0.05$ &   $0.38 \pm 0.01$ & $0.37 \pm 0.02$ & $4.844 \pm 0.002$ & $3391 \pm 157$ &    M3 V & 133 & PC  & - \\  
TIC 240968774 &          TOI-1467 &                 17,18,58 &   12.29 & 9.38 &   8.57 &  $37.52 \pm 0.03$ &  $0.49 \pm 0.01$ & $0.49 \pm 0.02$ & $4.746 \pm 0.008$ & $3834 \pm 157$ &      M2 V & 233 & PC  & - \\  
TIC 243185500 &          TOI-1468 &              17,42,43,57 &   12.50 & 9.34 &   8.50 &  $24.72 \pm 0.02$ &  $0.37 \pm 0.01$ & $0.36 \pm 0.02$ & $4.855 \pm 0.001$ & $3382 \pm 157$ &    M3.5 V  & 132 & 2$\times$CP  & 2022-10 \\  
TIC 201186294 &          TOI-1634 &                       18 &   13.22 & 9.48 &   8.60 &  $35.08 \pm 0.02$ &  $0.47 \pm 0.01$ & $0.46 \pm 0.02$ & $4.768 \pm 0.007$ & $3455 \pm 157$ &   M3 V & 131 & CP   & 2021-08 \\  
TIC 348673213 &          TOI-1639 &                    18,58 &   14.38 & 11.78 &  10.94 & $154.08 \pm 0.47$ &  $0.61 \pm 0.02$ & $0.60 \pm 0.02$ & $4.642 \pm 0.011$ & $3896 \pm 157$ &   M0.5 V & 150 & FP  & 2021-08 \\  
 TIC 28900646 &          TOI-1685 &                    19,59 &   13.38 & 9.62 &   8.76 &  $37.61 \pm 0.03$ &  $0.46 \pm 0.01$ & $0.46 \pm 0.02$ & $4.771 \pm 0.007$ & $3457 \pm 157$ &    M2 V & 248 & CP  & 2021-06 \\  
TIC 353475866 &          TOI-1693 &                19,43--45 &   12.96 & 9.19 &   8.33 &  $30.85 \pm 0.02$ &  $0.46 \pm 0.01$ & $0.46 \pm 0.02$ & $4.772 \pm 0.007$ & $3474 \pm 157$ &   M2.5 V & 260 & CP  & 2022-02 \\  
TIC 470381900 &          TOI-1696 &                    19,59 &   16.82 & 12.23 &  11.33 &  $64.60 \pm 0.12$ &  $0.28 \pm 0.01$ & $0.25 \pm 0.02$ & $4.948 \pm 0.008$ & $3181 \pm 157$ &   M4.5 V & 129 & CP & 2022-06 \\  
TIC 318022259 &          TOI-1730 &                 20,47,60 &   12.23 & 9.06 &   8.21 &  $35.65 \pm 0.03$ &  $0.53 \pm 0.02$ & $0.53 \pm 0.02$ & $4.709 \pm 0.009$ & $3691 \pm 157$ &   M0.5 V & 260 & 3$\times$PC  & - \\  
TIC 104208182 &          TOI-1738 &                    18,19 &   14.07 & 11.31 &  10.46 & $141.84 \pm 0.40$ &  $0.67 \pm 0.02$ & $0.64 \pm 0.02$ & $4.594 \pm 0.012$ & $3948 \pm 157$ &    M0 V & 189 & PC  & - \\  
TIC 408636441 &          TOI-1759 &          16,17,24,56--58 &   11.93 & 8.77 &   7.93 &  $40.13 \pm 0.02$ &  $0.63 \pm 0.02$ & $0.61 \pm 0.02$ & $4.629 \pm 0.011$ & $3960 \pm 157$ &    M2 V & 200 &  CP  & 2022-03 \\  
TIC 389900760 &          TOI-2120 &           18,24,25,52,58 &   15.65 & 11.10 &  10.21 &  $32.19 \pm 0.02$ &  $0.24 \pm 0.01$ & $0.21 \pm 0.02$ & $5.000 \pm 0.015$ & $3179 \pm 157$ &   M4.5 V & 164 & PC  & - \\  
TIC 329148988 &          TOI-2285 &           16,17,24,56,57 &   13.40 & 9.86 &   9.03 &  $42.48 \pm 0.04$ &  $0.46 \pm 0.01$ & $0.46 \pm 0.02$ & $4.773 \pm 0.007$ & $3546 \pm 157$ &    M2 V & 205 & CP  & 2022-02 \\  
TIC 321688498 &          TOI-2290 &        17,18,24,25,57,58 &   12.64 & 9.90 &   9.07 &  $58.11 \pm 0.03$ &  $0.57 \pm 0.02$ & $0.56 \pm 0.02$ & $4.680 \pm 0.010$ & $3863 \pm 157$ &    M0.5 V  & 147 & PC  & - \\  
TIC 168751223 &          TOI-2331 &                   4,5,31 &  15.08 & 12.13 &  11.28 & $162.07 \pm 0.53$ &   $0.57 \pm 0.02$ & $0.57 \pm 0.02$ & $4.674 \pm 0.011$ &  $3798 \pm 157$ &     M0.5V & 110 & PC & - \\  
TIC 212957629 &          TOI-2406 &               3,30,42,43 &   17.30 & 12.63 &  11.89 &  $55.62 \pm 0.12$ &  $0.20 \pm 0.01$ & $0.17 \pm 0.02$ & $5.070 \pm 0.026$ & $3163 \pm 157$ &     M4 V & 137 & CP  & 2021-09 \\  
TIC 350043412 &          TOI-5131 &                    45,46 & 14.28 & 12.03 &  11.27 & $216.92 \pm 0.94$ &  $0.63 \pm 0.03$ & $0.69 \pm 0.08$ & $4.674 \pm 0.069$ & $4427 \pm 122$ &        K3 V & 183 & PC  & - \\  
TIC 330637910 &          TOI-5561 &                 44,45,46 & 17.53 & 13.74 & 12.88 & $272.48 \pm 4.45$ &  $0.49 \pm 0.02$ & $0.49 \pm 0.02$ & $4.746 \pm 0.017$ & $3449 \pm 157$ &    M2.5 V & 99 & PC  & - \\
\enddata
\tablecomments{Columns from left to right are as follows: the EPIC or TIC ID of the star \citep{huber2014, TIC2018, TIC2019}; the alternative identifier of the star; the \textit{K2} or \textit{TESS} sectors in which the star was observed; the \textit{V} magnitude according to the TIC; the \textit{J} and \textit{Ks} magnitudes according to 2MASS \citep{2MASS2006}; {Gaia DR3 distance} of the star \cite{gaia2022}; the stellar radius, mass, surface gravity, and effective temperature according to the EPIC or TIC; the best-fit spectral type of the star from our spectrum (see Section \ref{sec: detailed_classification}); the maximum S/N of our spectrum of the star; a flag indicating the number and current status (as of the writing of this paper) of planet candidates around the star according to the NASA Exoplanet Archive and ExoFOP (where CP~=~Confirmed~Planet, PC~=~Planet~Candidate, and FP~=~False~Positive); {and the year and month that the planet candidates were determined to be confirmed planets (i.e., the publication date of the discovery paper) or false positives (which is listed on ExoFOP)}. {TOI-523 and TOI-5131 did not have uncertainties for $R_\star$ or $M_\star$ in the TIC. The uncertainties for those two targets were calculated manually using the relations defined in Section 2.3.5 of \citet{TIC2019}.}}
\end{deluxetable*}
\end{longrotatetable}

\section{Observations} \label{sec:observations}

Our observing procedure follows that used in similar previous studies \citep[e.g.,][]{dressing2017, dressing2019}. We observed our targets with the SpeX spectrograph on the 3-meter NASA IRTF. Our observations were acquired during the 2018B -- 2020B semesters under programs 2018B112 (PI: Dressing), 2019A072 (PI: Dressing), and 2020B115 (PI: Giacalone). We collected our observations in SXD mode with a $0\farcs3 \times 15\arcsec$ slit, providing a spectral resolution of $R \approx 2000$ over a wavelength range of $0.7 - 2.55 \, \mu{\rm m}$ \citep{2003PASP..115..362R, 2004SPIE.5492.1498R}. We experienced sub-arcsecond seeing on all nights of observation.

All of our observations were collected using an ABBA nod pattern, with the A and B positions located $7\farcs5$ apart, in order to remove telescope and sky backgrounds. We also synced the slit to the parallactic angle in all cases to ensure that atmospheric dispersion occurred along the slit (minimizing light loss) and to reduce systematics that could affect the measured spectral shapes of our targets. The ABBA nod pattern was repeated with the goal of achieving a S/N of at least 100 per resolution element. The measured S/N of each spectrum is listed in Table~\ref{tab: initial_params}.

In addition to our science targets, we acquired observations of A0V stars to use for telluric correction \citep{2003PASP..115..389V}. These A0V stars were observed within 1 hour of our science targets and had angular distances and airmass differences no greater than $15^\circ$ and 0.1, respectively.

Lastly, we obtained calibrations throughout the night using internal quartz and thorium-argon lamps. Because SpeX is physically attached to the telescope, we ran a new calibration sequence every time we moved to a new part of the sky to account for systematics stemming from telescope pointing.

We reduced our spectra using the \texttt{Spextool} reduction pipeline \citep{2004PASP..116..362C}. Telluric contamination in our spectra was removed using our data of A0V stars and \texttt{xtellcor} \citep{2003PASP..115..389V}, a tool included within \texttt{Spextool}.

\section{Data Analysis and Stellar Characterization} \label{sec:analysis}

\subsection{Initial Classification}

\begin{figure*}
    \centering
    \includegraphics[width=1.0\textwidth]{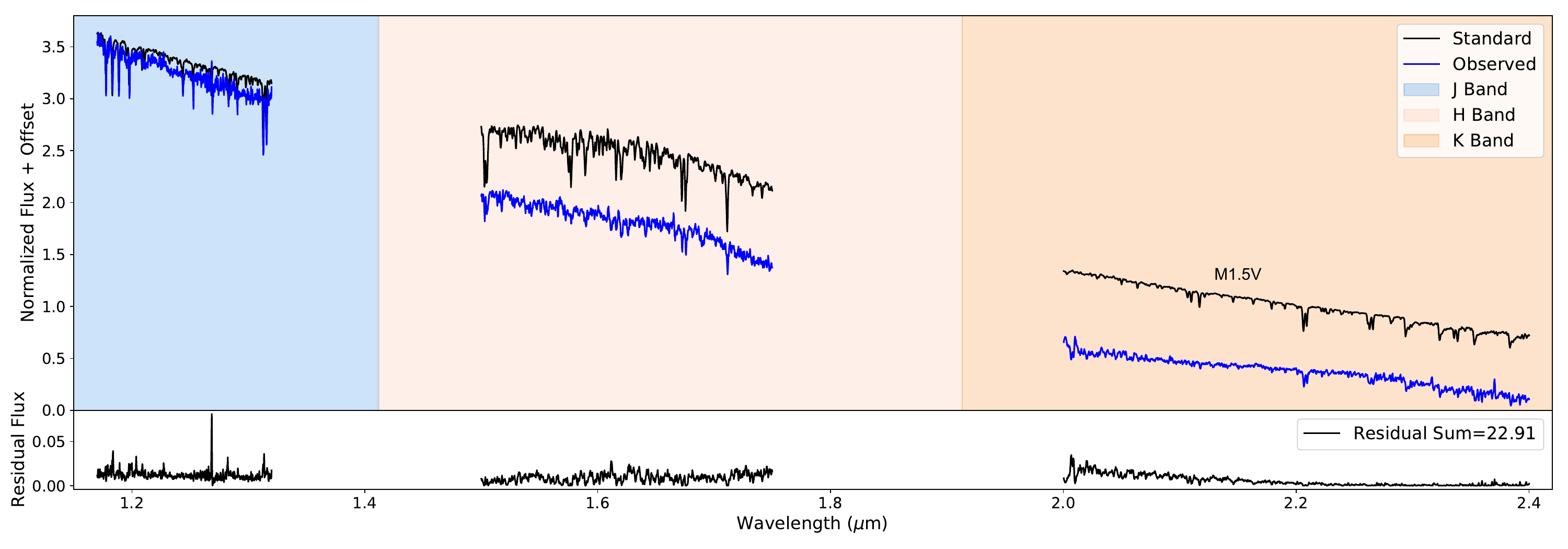}
    \includegraphics[width=1.0\textwidth]{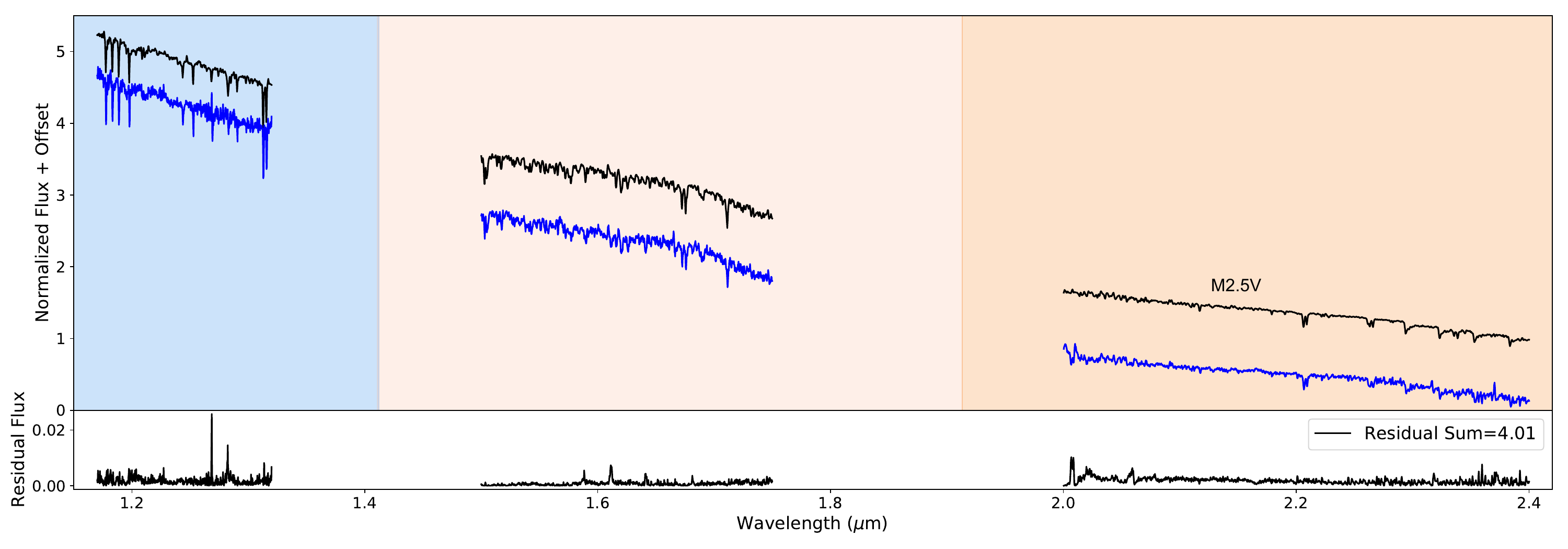}
    \includegraphics[width=1.0\textwidth]{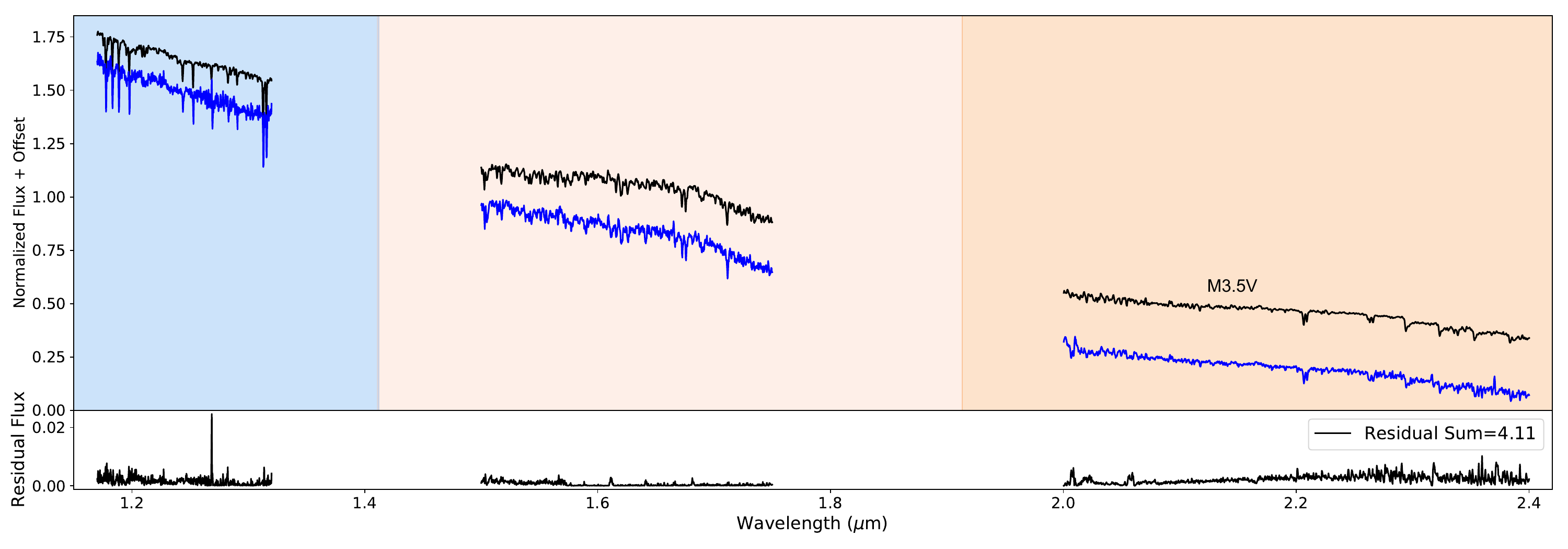}
    \caption{The observed spectrum for EPIC 212048748 compared with the IRTF standard spectrum (offset vertically for visual clarity) of HD 36395 (M1.5V; top) Gl 381 (M2.5V; middle), Gl 273 (M3.5V; bottom) within the highlighted J, H, and K bands. The bottom panels display the residuals between the observed and standard spectra, with Gl 381 providing the best match to the data. Data strongly influenced by telluric contamination has been masked out.}
    \label{fig: spec_comparison}
\end{figure*}

In a similar manner as \cite{dressing2017, dressing2019}, we started our analysis by performing an initial classification of our targets based on their spectra. For each star, we first split the spectrum into three segments corresponding to the $J$, $H$, and $Ks$ passbands and mean normalized each. We then did the same for all stars in the IRTF Spectral Library with M, K, and G spectral types \citep{2009ApJS..185..289R}. After correcting for differences in radial velocity, we calculated the $\chi^2$ between the target spectrum and each library spectrum in each passband separately, identifying the library spectrum that provides the best fit at each range of wavelengths. For stars where all passbands agreed on the best-fit library spectrum, we assigned the corresponding spectral type. For stars where different passbands preferred different spectral types, we visually inspected the full spectra to determine which library spectrum provided the best match. We believe this method to provide spectral types that are accurate to within one subtype. We show the results of this analysis in Table~\ref{tab: initial_params} and visualize the comparison in Figure~\ref{fig: spec_comparison}.

\subsection{Identifying Possible Stellar Binaries}

Before performing a detailed analysis of our targets to determine their metallicities and stellar properties, we sought to determine if any of our targets have nearby ($< 2 \arcsec$) stellar companions. Identifying these companions is important because their point spread functions are typically blended with the primary star, causing them to influence the spectra and apparent magnitudes of the sources. As a result, they can affect the stellar parameters that we extract from the data and lead to erroneous results (see Section \ref{sec: detailed_classification}).

First, we searched for known stellar companions by querying Gaia DR3 in a $2\arcsec$ cone around each star. For TIC 296780789, we identified the nearby star TIC 876200725 ($\Delta T = 2.10$, $\Delta G = 0.85$, ${\rm sep} = 1\farcs33$, ${\rm PA} = 123^\circ$ East of North).\footnote{Here, $T$ is {\it TESS} magnitude and $G$ is Gaia magnitude.} The two stars share similar parallaxes ($\pi = 11.27 \, {\rm mas}$ and $\pi = 11.23 \, {\rm mas}$, respectively) and proper motions (${\rm pmra} = -51.68 \, {\rm mas \, yr^{-1}}$ and ${\rm pmdec} = -92.24 \, {\rm mas \, yr^{-1}}$ for TIC 296780789,  ${\rm pmra} = -46.98 \, {\rm mas \, yr^{-1}}$ and ${\rm pmdec} = -95.18 \, {\rm mas \, yr^{-1}}$ for TIC 876200725), indicating that the two stars are likely gravitationally bound. In addition, for TIC 348673213, we identified the nearby star TIC 624816660 ($\Delta T = 4.83$, $\Delta G = 4.31$, ${\rm sep} = 1\farcs80$, ${\rm PA} = 184^\circ$ East of North). These two stars do not share close parallaxes or proper motions, meaning they are most likely chance-aligned. We note that TIC 296780789 is listed as having an ``ambiguous'' PC on ExoFOP, indicating that it will be difficult to determine if the planet is real due to the presence of the nearby star and has a relatively high chance of being an FP, and TIC 348673213 is listed as being an FP. Due to the presence of the blended stars, we remove both of these targets from further analysis.

Next, we searched for evidence of binary stars that are unresolved by Gaia photometry. To do this, we queried Gaia DR3 and determined the Renormalised Unit Weight Error (RUWE) for each star. RUWE is known as a useful metric for binarity, where stars with ${\rm RUWE} > 1.4$ are significantly more likely to be members of unresolved binary systems than those with smaller values \citep{lindegren2018, belokurov2020, elbadry2021}. We identified only two stars in our sample with high RUWE: TIC 365639282 ($\mathrm{RUWE} = 10.50$) and TIC 93125144 ($\mathrm{RUWE} = 12.56$). As we note in Table \ref{tab: initial_params}, the TOIs associated with both of these stars have already been designated as FPs. We removed these targets from further analysis. After this vetting step, our sample contained 61 stars.

\subsection{Detailed Classification}\label{sec: detailed_classification}

After our initial classification and removing probable binaries, we performed a more detailed analysis to extract metallicities from our spectra and calculate updated stellar parameters. We began by shifting our spectra into the lab frame using \texttt{tellrv}, a code that measures absolute radial velocities of stars using near-infrared spectra \citep{Newton2014, Newton2022}.\footnote{\url{https://github.com/ernewton/tellrv}} For the analysis in this paper, we converted the code from its native \texttt{IDL} to Python, which we have made available on GitHub.\footnote{\url{https://github.com/stevengiacalone/tellrv}} In short, the code works by wavelength-calibrating the spectrum of the target star using telluric features in the near infrared and then cross-correlating it with a standard star with a known radial velocity. All of our lab-frame spectra are publicly available and can be downloaded on ExoFOP.\footnote{\url{https://exofop.ipac.caltech.edu/tess/}} After this correction, we calculated the metallicities and updated stellar parameters for each of our targets using the following methods.

\subsubsection{Spectrum-Derived Stellar Metallicities}

Using our lab-frame spectra, we calculated metallicities ([Fe/H] and [M/H]) using the relations defined in \citet{Mann2013}, which are valid for cool dwarfs with spectral types between K7 and M5. As is shown in Table \ref{tab: initial_params}, some of the stars observed fall outside of this spectral type range according to our initial classification: EPIC 211586387 (K2 V), EPIC 211673349 (K4 V), EPIC 211731298 (K2 V), EPIC 211925595 (K2 V), EPIC 248527514 (K2-401; K5 V), TIC 101011575 (TOI-560; K2 V), and TIC 350043412 (TOI-5131; K3 V). We removed these stars from further analysis to avoid reporting erroneous metallicities and stellar parameters. The \citet{Mann2013} relations allow for the calculation of metallicity across several passbands, but \citet{dressing2019} reported metallicities derived using the $Ks$-band to be the most reliable in spectra collected with SpeX, likely due to a lower level of contamination from tellurics compared to the $H$-band. We therefore adopt values of [Fe/H] and [M/H] derived from the $Ks$-band spectra, adding a systematic uncertainty of 0.08 dex in quadrature with the uncertainties returned by the \citet{Mann2013} relations, as is recommended in the header of the code on GitHub.\footnote{\url{https://github.com/awmann/metal}} We report these metallicities in Table~\ref{tab: updated_params}.

\subsubsection{{\normalfont Gaia}-Derived Stellar Properties}

\startlongtable
\begin{deluxetable*}{lcccccc}\label{tab: updated_params}
\tabletypesize{\footnotesize}
\tablewidth{\textwidth}
 \tablecaption{Updated Stellar Parameters}
 \tablehead{
 \colhead{ID} & \colhead{$T_{\rm eff}$ (K)} & \colhead{$R_\star$ ($R_\odot$)} &  \colhead{$M_\star$ ($M_\odot$)} & \colhead{$L_\star$ ($L_\odot$)} & \colhead{[Fe/H] (dex)} & \colhead{[M/H] (dex)}
 }
\startdata
EPIC 211510580 & $3359 \pm 54$ & $0.356 \pm 0.010$ & $0.340 \pm 0.020$ & $0.0145 \pm 0.0004$ & $-0.044 \pm 0.093$ & $-0.078 \pm 0.090$ \\
EPIC 211663688 & $3036 \pm 54$ & $0.193 \pm 0.006$ & $0.160 \pm 0.020$ & $0.0028 \pm 0.0001$ & $-0.137 \pm 0.162$ & $-0.035 \pm 0.159$ \\
EPIC 211863149 & $3498 \pm 54$ & $0.451 \pm 0.013$ & $0.453 \pm 0.020$ & $0.0274 \pm 0.0007$ & $0.237 \pm 0.088$ & $0.146 \pm 0.087$ \\
EPIC 212048748 & $3289 \pm 50$ & $0.317 \pm 0.009$ & $0.293 \pm 0.020$ & $0.0106 \pm 0.0003$ & $-0.145 \pm 0.083$ & $-0.167 \pm 0.082$ \\
EPIC 212088059 & $3686 \pm 65$ & $0.531 \pm 0.015$ & $0.534 \pm 0.020$ & $0.0469 \pm 0.0019$ & $0.227 \pm 0.088$ & $0.177 \pm 0.085$ \\
EPIC 248433930 & $3786 \pm 58$ & $0.540 \pm 0.015$ & $0.536 \pm 0.020$ & $0.0540 \pm 0.0013$ & $-0.055 \pm 0.095$ & $-0.026 \pm 0.093$ \\
EPIC 248440276 & $3427 \pm 54$ & $0.404 \pm 0.011$ & $0.393 \pm 0.020$ & $0.0203 \pm 0.0006$ & $-0.161 \pm 0.085$ & $-0.140 \pm 0.084$ \\
EPIC 248731669 & $3700 \pm 72$ & $0.534 \pm 0.016$ & $0.532 \pm 0.021$ & $0.0482 \pm 0.0024$ & $-0.027 \pm 0.103$ & $-0.020 \pm 0.099$ \\
EPIC 248775938 & $3957 \pm 69$ & $0.659 \pm 0.019$ & $0.635 \pm 0.021$ & $0.0958 \pm 0.0036$ & $-0.008 \pm 0.093$ & $-0.037 \pm 0.091$ \\
EPIC 248856002 & $3592 \pm 60$ & $0.482 \pm 0.014$ & $0.478 \pm 0.021$ & $0.0349 \pm 0.0012$ & $-0.149 \pm 0.092$ & $-0.114 \pm 0.089$ \\
EPIC 248861279 & $3761 \pm 61$ & $0.556 \pm 0.016$ & $0.548 \pm 0.020$ & $0.0557 \pm 0.0019$ & $-0.153 \pm 0.087$ & $-0.116 \pm 0.085$ \\
TIC 2760710 & $3075 \pm 49$ & $0.243 \pm 0.007$ & $0.213 \pm 0.020$ & $0.0048 \pm 0.0001$ & $0.017 \pm 0.093$ & $-0.062 \pm 0.089$ \\
TIC 415969908 & $3416 \pm 53$ & $0.385 \pm 0.011$ & $0.369 \pm 0.020$ & $0.0181 \pm 0.0004$ & $-0.274 \pm 0.086$ & $-0.199 \pm 0.084$ \\
TIC 12423815 & $3744 \pm 65$ & $0.544 \pm 0.016$ & $0.550 \pm 0.021$ & $0.0523 \pm 0.0020$ & $0.409 \pm 0.106$ & $0.279 \pm 0.104$ \\
TIC 118327550 & $3398 \pm 55$ & $0.409 \pm 0.012$ & $0.400 \pm 0.021$ & $0.0201 \pm 0.0005$ & $-0.083 \pm 0.132$ & $-0.052 \pm 0.129$ \\
TIC 120916706 & $3402 \pm 74$ & $0.452 \pm 0.016$ & $0.447 \pm 0.022$ & $0.0247 \pm 0.0013$ & $-0.115 \pm 0.163$ & $-0.037 \pm 0.158$ \\
TIC 70899085 & $3903 \pm 61$ & $0.591 \pm 0.017$ & $0.593 \pm 0.020$ & $0.0730 \pm 0.0019$ & $0.469 \pm 0.084$ & $0.327 \pm 0.083$ \\
TIC 452866790 & $3357 \pm 52$ & $0.357 \pm 0.010$ & $0.339 \pm 0.020$ & $0.0145 \pm 0.0004$ & $-0.142 \pm 0.084$ & $-0.145 \pm 0.083$ \\
TIC 19025965 & $4340 \pm 70$ & $0.706 \pm 0.020$ & $0.673 \pm 0.020$ & $0.1591 \pm 0.0052$ & $0.144 \pm 0.084$ & $0.108 \pm 0.083$ \\
TIC 348538431 & $3507 \pm 57$ & $0.504 \pm 0.014$ & $0.500 \pm 0.020$ & $0.0346 \pm 0.0011$ & $-0.143 \pm 0.087$ & $-0.157 \pm 0.085$ \\
TIC 318937509 & $3402 \pm 56$ & $0.525 \pm 0.015$ & $0.509 \pm 0.020$ & $0.0333 \pm 0.0011$ & $-0.652 \pm 0.105$ & $-0.456 \pm 0.101$ \\
TIC 218795833 & $3287 \pm 55$ & $0.351 \pm 0.010$ & $0.339 \pm 0.020$ & $0.0130 \pm 0.0005$ & $0.210 \pm 0.102$ & $0.105 \pm 0.097$ \\
TIC 27649847 & $3472 \pm 53$ & $0.420 \pm 0.012$ & $0.414 \pm 0.020$ & $0.0231 \pm 0.0006$ & $-0.016 \pm 0.082$ & $-0.037 \pm 0.082$ \\
TIC 200593988 & $3557 \pm 56$ & $0.474 \pm 0.013$ & $0.464 \pm 0.020$ & $0.0324 \pm 0.0009$ & $-0.422 \pm 0.081$ & $-0.306 \pm 0.081$ \\
TIC 387690507 & $3735 \pm 61$ & $0.531 \pm 0.015$ & $0.538 \pm 0.020$ & $0.0495 \pm 0.0017$ & $0.393 \pm 0.096$ & $0.252 \pm 0.093$ \\
TIC 144700903 & $3932 \pm 61$ & $0.609 \pm 0.017$ & $0.607 \pm 0.020$ & $0.0800 \pm 0.0020$ & $0.459 \pm 0.092$ & $0.318 \pm 0.090$ \\
TIC 237751146 & $3402 \pm 59$ & $0.437 \pm 0.013$ & $0.432 \pm 0.021$ & $0.0230 \pm 0.0009$ & $-0.046 \pm 0.101$ & $-0.079 \pm 0.097$ \\
TIC 50618703 & $4144 \pm 64$ & $0.662 \pm 0.018$ & $0.635 \pm 0.020$ & $0.1166 \pm 0.0032$ & $-0.161 \pm 0.083$ & $-0.143 \pm 0.082$ \\
TIC 192826603 & $3840 \pm 65$ & $0.572 \pm 0.016$ & $0.560 \pm 0.020$ & $0.0640 \pm 0.0023$ & $-0.253 \pm 0.103$ & $-0.164 \pm 0.101$ \\
TIC 170849515 & $3877 \pm 77$ & $0.646 \pm 0.020$ & $0.622 \pm 0.021$ & $0.0849 \pm 0.0044$ & $-0.169 \pm 0.137$ & $0.001 \pm 0.132$ \\
TIC 55488511 & $3828 \pm 60$ & $0.576 \pm 0.016$ & $0.564 \pm 0.020$ & $0.0641 \pm 0.0017$ & $-0.244 \pm 0.086$ & $-0.229 \pm 0.084$ \\
TIC 413248763 & $3369 \pm 56$ & $0.363 \pm 0.010$ & $0.346 \pm 0.020$ & $0.0153 \pm 0.0006$ & $-0.187 \pm 0.082$ & $-0.161 \pm 0.081$ \\
TIC 1133072 & $3543 \pm 59$ & $0.488 \pm 0.014$ & $0.496 \pm 0.020$ & $0.0339 \pm 0.0012$ & $0.385 \pm 0.097$ & $0.318 \pm 0.095$ \\
TIC 296739893 & $3696 \pm 60$ & $0.540 \pm 0.015$ & $0.543 \pm 0.020$ & $0.0491 \pm 0.0015$ & $0.232 \pm 0.093$ & $0.176 \pm 0.091$ \\
TIC 32497972 & $3904 \pm 60$ & $0.600 \pm 0.017$ & $0.590 \pm 0.020$ & $0.0755 \pm 0.0018$ & $0.023 \pm 0.089$ & $0.018 \pm 0.087$ \\
TIC 29960110 & $3659 \pm 73$ & $0.481 \pm 0.014$ & $0.478 \pm 0.021$ & $0.0374 \pm 0.0021$ & $-0.091 \pm 0.082$ & $-0.121 \pm 0.082$ \\
TIC 163539739 & $3744 \pm 59$ & $0.553 \pm 0.015$ & $0.545 \pm 0.020$ & $0.0541 \pm 0.0017$ & $-0.182 \pm 0.086$ & $-0.146 \pm 0.085$ \\
TIC 13684720 & $3463 \pm 54$ & $0.473 \pm 0.013$ & $0.475 \pm 0.020$ & $0.0290 \pm 0.0008$ & $0.193 \pm 0.086$ & $0.097 \pm 0.084$ \\
TIC 343628284 & $3411 \pm 55$ & $0.379 \pm 0.011$ & $0.372 \pm 0.020$ & $0.0175 \pm 0.0005$ & $0.213 \pm 0.090$ & $0.133 \pm 0.088$ \\
TIC 240968774 & $3676 \pm 55$ & $0.498 \pm 0.014$ & $0.489 \pm 0.020$ & $0.0408 \pm 0.0008$ & $-0.334 \pm 0.083$ & $-0.295 \pm 0.082$ \\
TIC 243185500 & $3475 \pm 60$ & $0.369 \pm 0.010$ & $0.356 \pm 0.020$ & $0.0179 \pm 0.0007$ & $0.033 \pm 0.088$ & $-0.017 \pm 0.087$ \\
TIC 201186294 & $3517 \pm 53$ & $0.459 \pm 0.013$ & $0.461 \pm 0.020$ & $0.0291 \pm 0.0007$ & $0.161 \pm 0.090$ & $0.122 \pm 0.087$ \\
TIC 28900646 & $3533 \pm 54$ & $0.458 \pm 0.013$ & $0.460 \pm 0.020$ & $0.0295 \pm 0.0007$ & $0.158 \pm 0.082$ & $0.088 \pm 0.082$ \\
TIC 353475866 & $3529 \pm 56$ & $0.458 \pm 0.013$ & $0.459 \pm 0.020$ & $0.0293 \pm 0.0008$ & $0.154 \pm 0.082$ & $0.094 \pm 0.081$ \\
TIC 470381900 & $3198 \pm 97$ & $0.276 \pm 0.008$ & $0.253 \pm 0.020$ & $0.0072 \pm 0.0008$ & $0.333 \pm 0.088$ & $0.233 \pm 0.086$ \\
TIC 318022259 & $3699 \pm 58$ & $0.532 \pm 0.015$ & $0.528 \pm 0.020$ & $0.0478 \pm 0.0014$ & $-0.083 \pm 0.082$ & $-0.066 \pm 0.082$ \\
TIC 104208182 & $4041 \pm 61$ & $0.658 \pm 0.018$ & $0.641 \pm 0.020$ & $0.1041 \pm 0.0026$ & $0.252 \pm 0.085$ & $0.204 \pm 0.084$ \\
TIC 408636441 & $3890 \pm 70$ & $0.622 \pm 0.017$ & $0.610 \pm 0.020$ & $0.0799 \pm 0.0036$ & $0.108 \pm 0.085$ & $0.157 \pm 0.084$ \\
TIC 389900760 & $3138 \pm 50$ & $0.240 \pm 0.007$ & $0.211 \pm 0.020$ & $0.0050 \pm 0.0001$ & $0.128 \pm 0.085$ & $0.047 \pm 0.084$ \\
TIC 329148988 & $3574 \pm 57$ & $0.458 \pm 0.013$ & $0.458 \pm 0.020$ & $0.0309 \pm 0.0010$ & $0.072 \pm 0.083$ & $0.034 \pm 0.083$ \\
TIC 321688498 & $3851 \pm 61$ & $0.566 \pm 0.016$ & $0.560 \pm 0.020$ & $0.0634 \pm 0.0017$ & $-0.033 \pm 0.088$ & $-0.029 \pm 0.086$ \\
TIC 168751223 & $3832 \pm 65$ & $0.562 \pm 0.016$ & $0.562 \pm 0.021$ & $0.0614 \pm 0.0021$ & $0.218 \pm 0.096$ & $0.094 \pm 0.095$ \\
TIC 212957629 & $3123 \pm 50$ & $0.203 \pm 0.006$ & $0.166 \pm 0.020$ & $0.0035 \pm 0.0001$ & $-0.492 \pm 0.088$ & $-0.392 \pm 0.087$ \\
TIC 330637910 & $3548 \pm 70$ & $0.490 \pm 0.015$ & $0.488 \pm 0.021$ & $0.0343 \pm 0.0017$ & $-0.050 \pm 0.102$ & $-0.017 \pm 0.100$ \\
\enddata
\end{deluxetable*}

We calculated updated stellar parameters using a combination of methods. First, we calculated the absolute $J$ and $Ks$ magnitudes ($M_J$ and $M_{Ks}$, respectively) of each star using stellar parallaxes reported in Gaia DR3 \citep{gaia2022}, propagating uncertainties based on those reported by 2MASS and Gaia. Next, we used the metallicity-dependent relation in Table 3 of \citet{Mann2015} to calculate the $V-J$ bolometric correction. We calculated uncertainties for the bolometric correction using the uncertainties in $V$ and $J$ reported in the TIC and 2MASS and the uncertainties in [Fe/H] calculated above, subsequently adding a systematic uncertainty of 0.012 in quadrature as recommended in \citet{Mann2015}. We then used the correction and $M_J$ to calculate the absolute bolometric magnitude and bolometric luminosity ($L_\star$) of each star. Next, we calculated values of $R_\star$ following the $R_\star - M_{Ks} - {\rm [Fe/H]}$ relation in Equation 5 and Table 1 of \citet{Mann2015}. We calculated uncertainties in $R_\star$ using those previously calculated for $M_{Ks}$ and [Fe/H], subsequently adding a systematic fractional uncertainty of $2.7 \%$ in quadrature. We continued by calculated values of $M_\star$ using the fifth-order $M_\star - M_{Ks}$ relation defined in Table 6 of \citet{Mann2019}. We calculated uncertainties in $M_\star$ using those previously calculated for $M_{Ks}$, subsequently adding a systematic uncertainty of $0.02 \, M_\odot$ in quadrature as recommended in \citet{Mann2019}. Similar to \citet{dressing2019}, we opted to not use the metallicity-dependent relations in \citet{Mann2019} to calculate $M_\star$, due to the current small sample of M dwarfs with precisely constrained masses. Lastly, we calculated $T_{\rm eff}$ using the Stefan-Boltzman relation. Our full set of updated parameters is shown in Table \ref{tab: updated_params}. For illustrative purposes, we show in Figure \ref{fig: HR} the locations of our targets on the Hertzsprung-Russell diagram using our updated stellar parameters. All stars in our sample lie within the main-sequence track, supporting the notion that they host bona fide planets rather than FPs caused by blended companion stars. Our updated parameters are compared to their respective catalog parameters in Figure \ref{fig: cat_v_derived}, demonstrating that our stellar properties provide a significant improvement over those reported in the EPIC\footnote{Although we acknowledge that updated catalogs have been published since the original release of the EPIC \citep[e.g.,][]{hardegree-ullman2020}.} and are generally consistent with (although often more precise than) those in the TIC.

 and bolometric luminosity ($L_\star$) of each star

\begin{figure}
    \centering
    \includegraphics[width=0.48\textwidth]{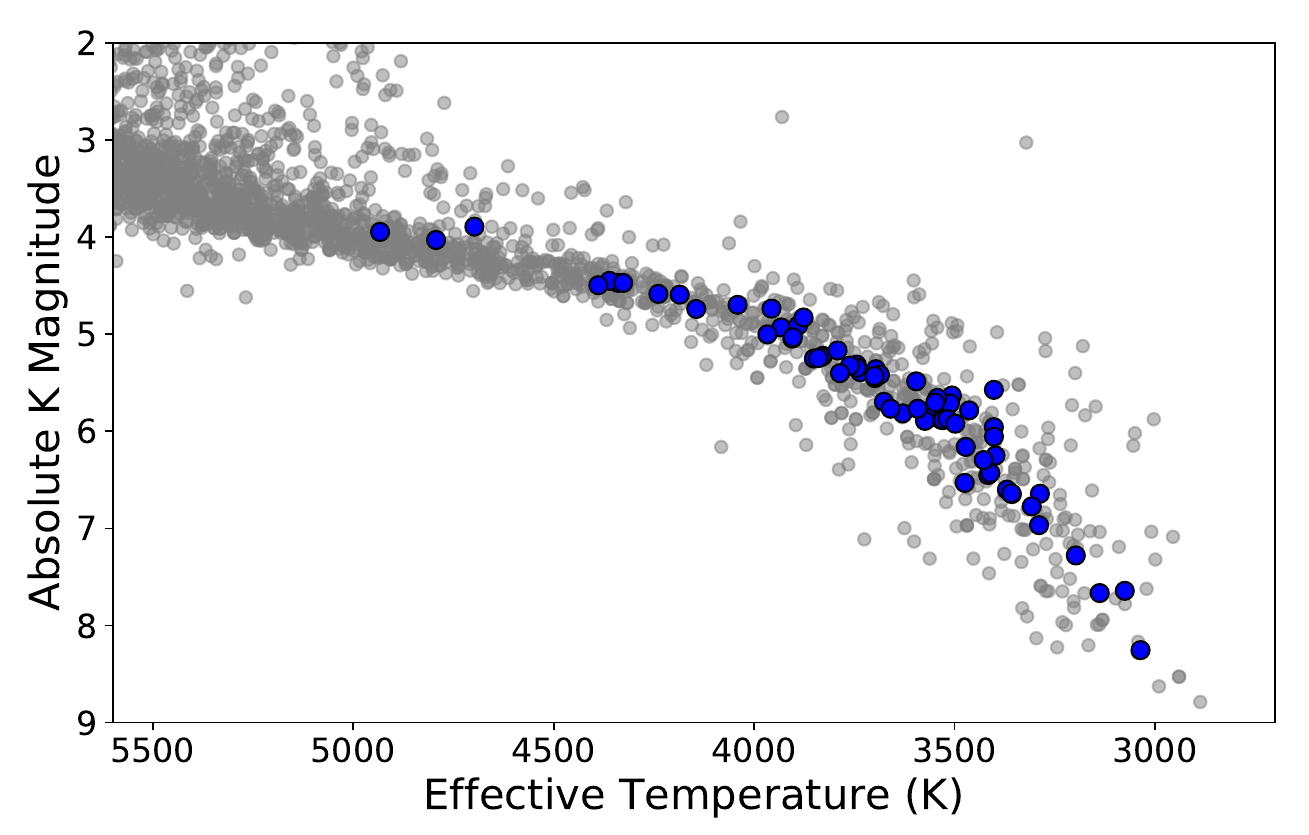}
    \caption{Hurtzsprung-Russell diagram using the updated values of our observed stars from Section \ref{sec:analysis} (blue circles). For reference, all {\it TESS} Objects of Interest are plotted in the background (grey circles).}
    \label{fig: HR}
\end{figure}

\begin{figure*}
    \centering
    \includegraphics[width=0.32\textwidth]{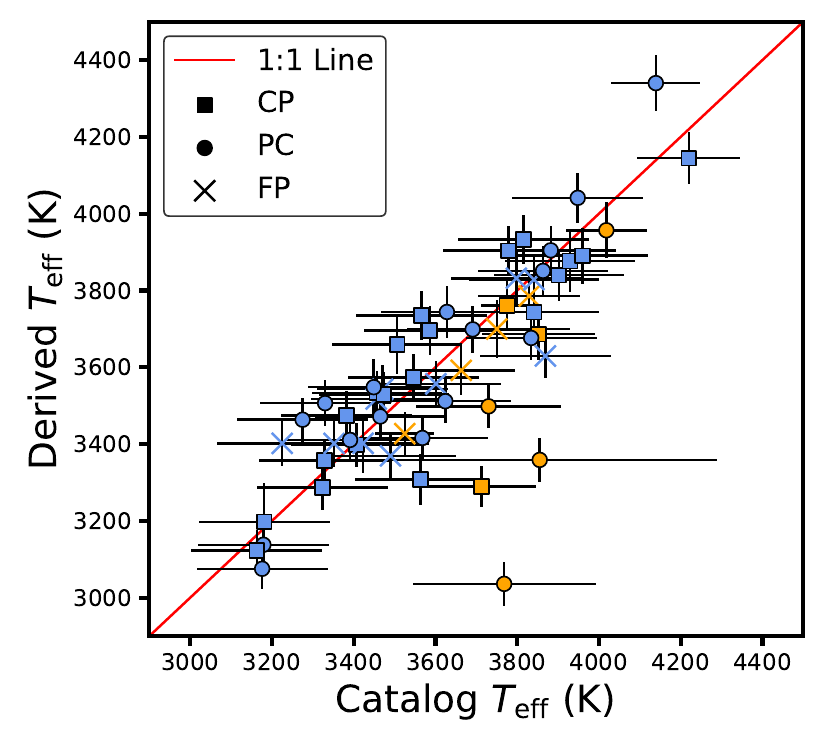}
    \includegraphics[width=0.32\textwidth]{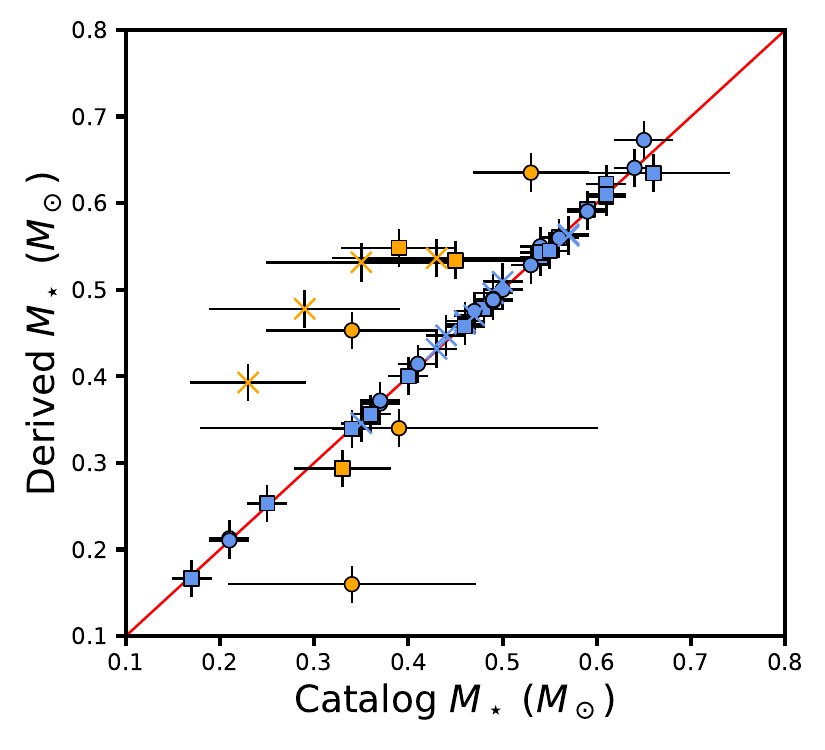}
    \includegraphics[width=0.32\textwidth]{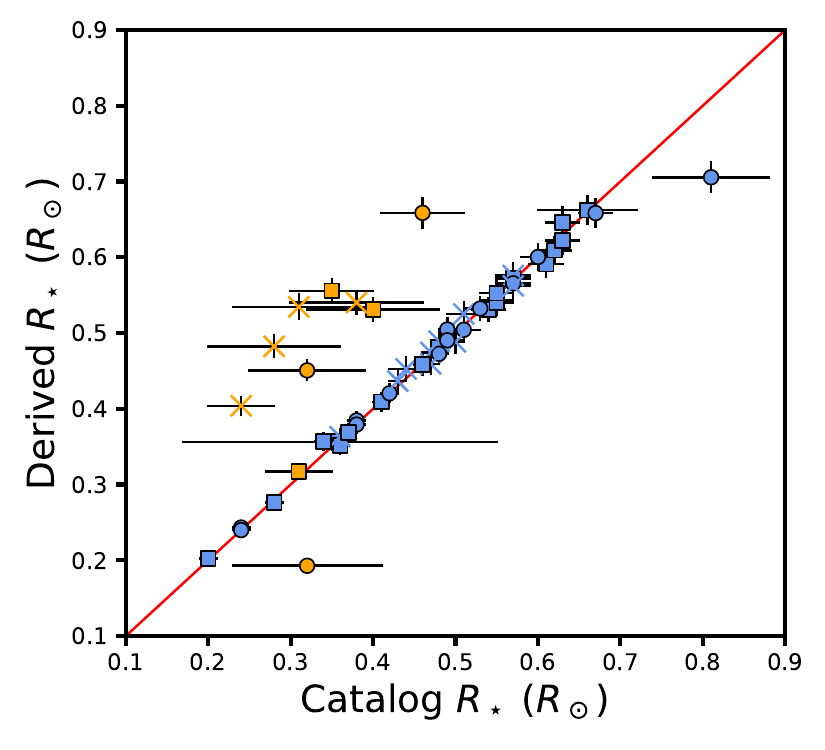}
    \caption{Derived stellar parameters, as discussed in Section \ref{sec:analysis}, plotted against the respective values listed in the EPIC (orange) and the TIC (light blue). Squares indicated stars with at least one confirmed planet, circles indicate stars with only planet candidates, and crosses indicate stars with only false positives or no planet candidates. The solid red line in each panel denotes the 1-to-1 line. In general, our values agree well with the values in the TIC and likely provide improved accuracy over those in the EPIC using stellar models. In many cases, we provide parameters with a greater precision than either catalog.}
    \label{fig: cat_v_derived}
\end{figure*}

\section{Updated Planet Parameters} \label{sec:planets}

Using our new stellar parameters, we performed light curve fits for all of the stars in our sample with confirmed planets and planet candidates to calculate updated planet radii and orbital properties. All targets listed as CP and PC in Table \ref{tab: initial_params} are included in these calculations with the exception of those removed from the analysis in the previous section. We also exclude EPIC 211663688 and EPIC 211863149 \citep{Yu2018}, as we were unable to recover the reported transits of their planet candidates in the \textit{K2} light curves.\footnote{\citet{Yu2018} reported these as planet candidates with orbital periods of 1.86 days and 2.61 days and with transit depths of $1.46 \%$ and $0.28 \%$, respectively. However, we note that these candidates were identified in uncalibrated pixels from the \textit{K2} mission, meaning they have a higher likelihood of being false positives caused by systematics or background contamination \citep[e.g.,][]{lehmann2024K2}.} Our new best-fit planet parameters are displayed in Tables \ref{tab: planet_params} and \ref{tab: planet_candidate_params}. We describe our calculation procedure below.

\subsection{Light Curve Generation}

We obtained the light curves for the confirmed planets in our sample using the \texttt{Lightkurve} Python package \citep{lightkurve2018}. For \textit{K2} targets, we used the light curves extracted with the EVEREST pipeline \citep{luger2016everest,luger2018everest}, which we found to produce light curves with the least noise and systematics for the stars in our sample after a visual comparison. For \textit{TESS} targets, we used the light curves extracted with the SPOC pipeline \citep{jenkins2016} and Quick Look Pipeline \citep{huang2020a, huang2020b}, with a preference for SPOC data when available due to the shorter cadence of the data and its more careful treatment of spacecraft systematics. Both \textit{K2} (\dataset[doi:10.17909/T9WS3R]{https://archive.stsci.edu/missions/k2/download_scripts/lightcurves/all/}) and \textit{TESS} (\dataset[doi:10.17909/t9-nmc8-f686]{https://archive.stsci.edu/tess/bulk_downloads/bulk_downloads_ffi-tp-lc-dv.html\#lc}) light curves can be downloaded on MAST \citep{MAST_K2, MAST_TESS}. For all light curves, we masked out data flagged as having poor quality and median-normalized the flux. We then flattened the light curves using the robust spline-fitting Huber-estimator algorithm as implemented in the \texttt{w{\={o}}tan} Python package \citep{hippke2019}, enforcing a window length (i.e., distance between knots) of 0.5 days. After visually inspecting the light curves using various knot distances, we determined that this choice does not distort the transit morphologies of the planets and planet candidates analyzed here, which all have transit durations under 3 hours.

\subsection{Light Curve Fits}

We performed our fits using the \texttt{exoplanet} Python package \citep{exoplanet2021}, which implements the Markov-Chain Monte Carlo capabilities of \texttt{pymc3} for transiting planet parameter estimation \citep{Salvatier2016}. For each planet, we began by determining the passband-appropriate quadratic limb darkening coefficients based on their updated values of $\log g$, $T_{\rm eff}$, and metallicity \citep{claret2017}. We initialized each model with the following priors: Gaussian priors on $M_\star$, $R_\star$, and $T_{\rm eff}$ based on the values in Table \ref{tab: updated_params}; Gaussian priors on the log of $P_{\rm orb}$, the log of the planet-star radius ratio ($R_{\rm p}/R_\star$), and the transit midpoint time ($T_0$), all centered on the values reported in the literature or on ExoFOP; a uniform prior on the transit impact parameter ($b$) that ranges from 0 to $1 + R_{\rm p}/R_\star$; and a prior on orbital eccentricity ($e$) based on \citet{kipping2013}. For each planet and planet candidate, we ran a 10 walker ensemble with 20,000 steps and discarded the first 10,000 steps as burn-in, ensuring convergence using the Gelman-Rubin statistic \citep{gelman-rubin1992}. The resulting best-fit parameters are shown in Tables \ref{tab: planet_params} and \ref{tab: planet_candidate_params}.

\begin{deluxetable*}{lcccccc}\label{tab: planet_params}

\tabletypesize{\footnotesize}
\tablewidth{\textwidth}
 \tablecaption{Best-Fit Planet Parameters for Confirmed Planets}
 \tablehead{
 \colhead{Name} & \colhead{$P_\mathrm{orb}$ (days)} & \colhead{$T_\mathrm{0}$ (KBJD or TBJD)\tablenotemark{a}} &  \colhead{$b$} & \colhead{$a$ (au)} & \colhead{$R_\mathrm{p}$ ($R_\oplus$)} & \colhead{$R_\mathrm{p, Lit}$ ($R_\oplus$)}
 }
\startdata
\vspace{3pt}
K2-313 b & $5.745909^{+0.000035}_{-0.000035}$ & $3664.7714^{+0.0022}_{-0.0022}$ & $0.58^{+0.11}_{-0.12}$ & $0.04170^{+0.00092}_{-0.00091}$ & $2.005^{+0.073}_{-0.074}$ & $2.025^{+0.096}_{-0.097}$ \\ \vspace{3pt}
K2-345 b & $10.367443^{+0.000030}_{-0.000030}$ & $2308.7110^{+0.0025}_{-0.0025}$ & $0.54^{+0.22}_{-0.27}$ & $0.07553^{+0.00094}_{-0.00097}$ & $1.99^{+0.12}_{-0.13}$ & $2.110^{+0.200}_{-0.009}$ \\ \vspace{3pt}
K2-404 b & $13.11372^{+0.00050}_{-0.00050}$ & $3084.8388^{+0.0014}_{-0.0014}$ & $0.44^{+0.20}_{-0.23}$ & $0.0891^{+0.0011}_{-0.0011}$ & $2.53^{+0.11}_{-0.11}$ & $2.530^{+0.084}_{-0.083}$ \\ \vspace{3pt}
TOI-244 b & $7.397216^{+0.000036}_{-0.000036}$ & $1357.3625^{+0.0028}_{-0.0028}$ & $0.83^{+0.16}_{-0.19}$ & $0.05477^{+0.00096}_{-0.00095}$ & $1.40^{+0.14}_{-0.17}$ & $1.52^{+0.12}_{-0.12}$ \\ \vspace{3pt}
TOI-442 b & $4.0520291^{+0.0000030}_{-0.0000031}$ & $2196.11444^{+0.00048}_{-0.00048}$ & $0.73^{+0.11}_{-0.11}$ & $0.04178^{+0.00048}_{-0.00048}$ & $4.66^{+0.23}_{-0.24}$ & $4.70^{+0.30}_{-0.30}$ \\ \vspace{3pt}
TOI-488 b & $1.1980059^{+0.0000019}_{-0.0000020}$ & $2228.97790^{+0.00086}_{-0.00086}$ & $0.25^{+0.19}_{-0.19}$ & $0.01538^{+0.00030}_{-0.00030}$ & $1.216^{+0.056}_{-0.056}$ & $1.264^{+0.050}_{-0.049}$ \\ \vspace{3pt}
TOI-519 b & $1.26523285^{+0.00000048}_{-0.00000048}$ & $1491.87690^{+0.00023}_{-0.00023}$ & $0.22^{+0.11}_{-0.12}$ & $0.01599^{+0.00032}_{-0.00032}$ & $11.70^{+0.34}_{-0.35}$ & $11.50^{+0.30}_{-0.30}$ \\ \vspace{3pt}
TOI-530 b & $6.3875854^{+0.0000082}_{-0.0000084}$ & $2530.54083^{+0.00075}_{-0.00077}$ & $0.16^{+0.13}_{-0.13}$ & $0.05476^{+0.00068}_{-0.00069}$ & $8.27^{+0.24}_{-0.24}$ & $9.30^{+0.70}_{-0.70}$ \\ \vspace{3pt}
TOI-532 b & $2.32684^{+0.00023}_{-0.00023}$ & $1470.5758^{+0.0012}_{-0.0012}$ & $0.30^{+0.18}_{-0.20}$ & $0.02911^{+0.00033}_{-0.00033}$ & $5.41^{+0.24}_{-0.24}$ & $5.82^{+0.19}_{-0.19}$ \\ \vspace{3pt}
TOI-544 b & $1.5483512^{+0.0000019}_{-0.0000019}$ & $1469.75771^{+0.00076}_{-0.00075}$ & $0.64^{+0.23}_{-0.31}$ & $0.02251^{+0.00024}_{-0.00023}$ & $2.14^{+0.14}_{-0.14}$ & $2.03^{+0.10}_{-0.10}$ \\ \vspace{3pt}
TOI-551 b & $2.6473058^{+0.0000025}_{-0.0000025}$ & $2203.15209^{+0.00062}_{-0.00062}$ & $1.03^{+0.11}_{-0.10}$ & $0.03089^{+0.00037}_{-0.00037}$ & $15.9^{+5.1}_{-4.8}$ & $14.9^{+6.8}_{-3.7}$ \\ \vspace{3pt}
TOI-555 b & $1.9416425^{+0.0000032}_{-0.0000032}$ & $1438.1464^{+0.0010}_{-0.0011}$ & $0.24^{+0.16}_{-0.16}$ & $0.02600^{+0.00029}_{-0.00029}$ & $11.90^{+0.42}_{-0.42}$ & $12.09^{+0.35}_{-0.35}$ \\ \vspace{3pt}
TOI-562 b & $3.9305992^{+0.0000042}_{-0.0000042}$ & $2272.67531^{+0.00048}_{-0.00049}$ & $0.43^{+0.20}_{-0.24}$ & $0.03424^{+0.00067}_{-0.00067}$ & $1.233^{+0.047}_{-0.048}$ & $1.217^{+0.084}_{-0.083}$ \\ \vspace{3pt}
TOI-620 b & $5.0988429^{+0.0000055}_{-0.0000056}$ & $1518.00433^{+0.00064}_{-0.00065}$ & $0.65^{+0.24}_{-0.29}$ & $0.04732^{+0.00058}_{-0.00058}$ & $3.08^{+0.21}_{-0.20}$ & $3.76^{+0.15}_{-0.15}$ \\ \vspace{3pt}
TOI-1201 b & $2.4919817^{+0.0000046}_{-0.0000046}$ & $1411.6698^{+0.0011}_{-0.0011}$ & $0.31^{+0.19}_{-0.20}$ & $0.02814^{+0.00040}_{-0.00040}$ & $2.183^{+0.087}_{-0.088}$ & $2.415^{+0.091}_{-0.090}$ \\ \vspace{3pt}
TOI-1278 b & $14.4779^{+0.0024}_{-0.0024}$ & $1711.9586^{+0.0018}_{-0.0019}$ & $1.030^{+0.086}_{-0.082}$ & $0.0950^{+0.0012}_{-0.0012}$ & $11.4^{+3.9}_{-3.7}$ & $12.2^{+2.7}_{-2.2}$ \\ \vspace{3pt}
TOI-1468 b & $1.8805192^{+0.0000046}_{-0.0000047}$ & $2448.30598^{+0.00055}_{-0.00058}$ & $0.30^{+0.19}_{-0.20}$ & $0.02116^{+0.00039}_{-0.00040}$ & $1.398^{+0.056}_{-0.056}$ & $1.280^{+0.038}_{-0.039}$ \\ \vspace{3pt}
TOI-1468 c & $15.532507^{+0.000055}_{-0.000057}$ & $2450.35637^{+0.00067}_{-0.00069}$ & $0.47^{+0.23}_{-0.27}$ & $0.0865^{+0.0016}_{-0.0016}$ & $2.005^{+0.089}_{-0.089}$ & $2.064^{+0.044}_{-0.044}$ \\ \vspace{3pt}
TOI-1634 b & $0.98941^{+0.00011}_{-0.00011}$ & $1791.5145^{+0.0016}_{-0.0016}$ & $0.30^{+0.19}_{-0.20}$ & $0.01503^{+0.00020}_{-0.00021}$ & $1.849^{+0.090}_{-0.090}$ & $1.790^{+0.080}_{-0.081}$ \\ \vspace{3pt}
TOI-1685 b & $0.66916^{+0.00011}_{-0.00011}$ & $1816.2247^{+0.0023}_{-0.0024}$ & $0.43^{+0.23}_{-0.26}$ & $0.01155^{+0.00016}_{-0.00017}$ & $1.54^{+0.10}_{-0.10}$ & $1.700^{+0.070}_{-0.070}$ \\ \vspace{3pt}
TOI-1693 b & $1.7666891^{+0.0000048}_{-0.0000050}$ & $2549.09556^{+0.00098}_{-0.00099}$ & $0.24^{+0.20}_{-0.19}$ & $0.02205^{+0.00031}_{-0.00032}$ & $1.371^{+0.057}_{-0.059}$ & $1.41^{+0.10}_{-0.10}$  \\ \vspace{3pt}
TOI-1696 b & $2.50052^{+0.00063}_{-0.00063}$ & $1816.6980^{+0.0036}_{-0.0036}$ & $0.38^{+0.23}_{-0.25}$ & $0.02283^{+0.00060}_{-0.00060}$ & $3.22^{+0.27}_{-0.29}$ & $3.09^{+0.11}_{-0.11}$ \\ \vspace{3pt}
TOI-1759 b & $18.85006^{+0.00018}_{-0.00018}$ & $2838.7691^{+0.0097}_{-0.0097}$ & $0.31^{+0.20}_{-0.21}$ & $0.1175^{+0.0013}_{-0.0013}$ & $3.33^{+0.13}_{-0.13}$ & $3.14^{+0.10}_{-0.10}$ \\ \vspace{3pt}
TOI-2285 b & $27.26957^{+0.00097}_{-0.00100}$ & $1747.1830^{+0.0051}_{-0.0049}$ & $0.58^{+0.23}_{-0.28}$ & $0.1367^{+0.0020}_{-0.0020}$ & $1.99^{+0.18}_{-0.19}$ & $1.740^{+0.080}_{-0.080}$ \\ \vspace{3pt}
TOI-2406 b & $3.076673^{+0.000015}_{-0.000015}$ & $2475.9470^{+0.0012}_{-0.0012}$ & $0.30^{+0.24}_{-0.24}$ & $0.02253^{+0.00093}_{-0.00095}$ & $2.88^{+0.11}_{-0.16}$ & $2.94^{+0.17}_{-0.16}$ \\
\enddata
\tablecomments{Updated values are medians and $68\%$ confidence intervals of the posterior distributions. The right-most column contains literature $R_\mathrm{p}$ values reported in the following papers: K2-131~b (G~9-40~b; \citealt{Stefansson2020}), K2-345~b \citep{deLeon2021}, K2-404~b \citep{Zink2021, Christiansen2022}, TOI-244~b \citep{Castro2023}, TOI-442~b (LP~714-47~b; \citealt{Dreizler2020}), TOI-488~b (GJ~3473~b; \citealt{Kemmer2020}), TOI-519~b (\citealt{Kagetani2023}, but also see \citealt{Parviainen2021}), TOI-530~b \citep{gan2022}, TOI-532~b \citep{Kanodia2021}, TOI-544~b \citep{Giacalone2022}, TOI-551~b (NGTS-1~b; \citealt{Bayliss2018}), TOI-555~b (HATS-76~b; \citealt{Jordan2022}), TOI-562~b (GJ~357~b; \citealt{Luque2019}), TOI-620~b \citep{Reefe2022}, TOI-1201~b \citep{Kossakowski2021}, TOI-1278~b \citep{Artigau2021}, TOI-1468~b and c \citep{Charturvedi2022}, TOI-1634~b (\citealt{Cloutier2021}, but also see \citealt{Hirano2021}), TOI-1685~b (\citealt{Bluhm2021}, but also see \citealt{Hirano2021}), TOI-1693~b \citep{Giacalone2022}, TOI-1696~b (\citealt{Mori2022}, but also see \citealt{Beard2022}), TOI-1759~b (\citealt{Espinoza2022}, but also see \citealt{Martioli2022}), TOI-2285~b \citep{Fukui2021}, and TOI-2406~b \citep{Wells2021}.}
\tablenotetext{a}{KBJD applies to {\it K2} targets and is BJD - 2454833. TBJD applies to {\it TESS} targets and is BJD - 2457000.}
\end{deluxetable*}

\begin{deluxetable*}{lcccccc}\label{tab: planet_candidate_params}

\tabletypesize{\footnotesize}
\tablewidth{\textwidth}
 \tablecaption{Best-Fit Planet Parameters for Planet Candidates}
 \tablehead{
 \colhead{Name} & \colhead{$P_\mathrm{orb}$ (days)} & \colhead{$T_\mathrm{0}$ (KBJD or TBJD)\tablenotemark{a}} &  \colhead{$b$} & \colhead{$a$ (au)} & \colhead{$R_\mathrm{p}$ ($R_\oplus$)} & \colhead{$R_\mathrm{p,PC}$ ($R_\oplus$)}
 }
\startdata
\vspace{3pt}
EPIC 211510580.01 & $5.31214^{+0.00017}_{-0.00018}$ & $2310.1441^{+0.0034}_{-0.0034}$ & $0.37^{+0.25}_{-0.26}$ & $0.04157^{+0.00083}_{-0.00083}$ & $1.32^{+0.11}_{-0.11}$ & $1.45^{+0.15}_{-0.96}$ \\ \vspace{3pt}   
EPIC 248775938.01 & $1.753837^{+0.000017}_{-0.000017}$ & $3074.84941^{+0.00087}_{-0.00089}$ & $0.894^{+0.079}_{-0.042}$ & $0.02447^{+0.00028}_{-0.00027}$ & $7.0^{+1.4}_{-1.1}$ & $5.06^{+0.49}_{-0.44}$ \\ \vspace{3pt}
TOI-233.01 & $11.670015^{+0.000038}_{-0.000037}$ & $1365.2609^{+0.0023}_{-0.0023}$ & $0.36^{+0.20}_{-0.22}$ & $0.0722^{+0.0013}_{-0.0013}$ & $2.058^{+0.094}_{-0.093}$ & $1.7^{+1.3}_{-1.3}$ \\ \vspace{3pt}   
TOI-233.02 & $7.201429^{+0.000050}_{-0.000054}$ & $1359.4673^{+0.0032}_{-0.0031}$ & $0.56^{+0.23}_{-0.28}$ & $0.05236^{+0.00094}_{-0.00095}$ & $1.73^{+0.12}_{-0.13}$ & $2.02^{+0.71}_{-0.71}$ \\ \vspace{3pt}  
TOI-234.01 & $2.8392834^{+0.0000059}_{-0.0000058}$ & $1356.0879^{+0.0012}_{-0.0012}$ & $0.73^{+0.22}_{-0.25}$ & $0.03215^{+0.00039}_{-0.00039}$ & $11.6^{+2.5}_{-1.9}$ & $11.15^{+0.50}_{-0.50}$ \\ \vspace{3pt} 
TOI-493.01 & $5.944436^{+0.000011}_{-0.000011}$ & $2575.21361^{+0.00083}_{-0.00082}$ & $0.51^{+0.20}_{-0.25}$ & $0.05628^{+0.00056}_{-0.00055}$ & $3.78^{+0.18}_{-0.18}$ & $4.14^{+0.38}_{-0.38}$ \\ \vspace{3pt}
TOI-507.01 & $0.89946587^{+0.00000020}_{-0.00000019}$ & $2963.62275^{+0.00082}_{-0.00079}$ & $0.94^{+0.13}_{-0.11}$ & $0.01446^{+0.00020}_{-0.00020}$ & $18.4^{+4.7}_{-4.2}$ & $17.1^{+3.8}_{-3.8}$ \\ \vspace{3pt}
TOI-521.01 & $1.5428486^{+0.0000031}_{-0.0000032}$ & $2575.7957^{+0.0010}_{-0.0010}$ & $0.28^{+0.22}_{-0.22}$ & $0.01947^{+0.00030}_{-0.00031}$ & $2.000^{+0.089}_{-0.088}$ & $1.99^{+0.12}_{-0.12}$ \\ \vspace{3pt}
TOI-557.01 & $3.344925^{+0.000016}_{-0.000016}$ & $1439.5474^{+0.0023}_{-0.0023}$ & $0.66^{+0.21}_{-0.28}$ & $0.03617^{+0.00043}_{-0.00043}$ & $2.35^{+0.17}_{-0.18}$ & $2.3^{+1.2}_{-1.2}$ \\ \vspace{3pt}
TOI-876.01 & $29.48287^{+0.00027}_{-0.00027}$ & $1457.3967^{+0.0051}_{-0.0049}$ & $0.75^{+0.18}_{-0.23}$ & $0.1567^{+0.0018}_{-0.0018}$ & $2.99^{+0.26}_{-0.26}$ & $2.8^{+1.4}_{-1.4}$ \\ \vspace{3pt}
TOI-1433.01 & $12.4379635^{+0.0000066}_{-0.0000064}$ & $1695.01136^{+0.00044}_{-0.00080}$ & $0.98^{+0.23}_{-0.20}$ & $0.0820^{+0.0011}_{-0.0011}$ & $27.2^{+8.8}_{-7.5}$ & $24^{+46}_{-24}$ \\ \vspace{3pt}
TOI-1448.01 & $8.1135^{+0.0020}_{-0.0020}$ & $1713.3338^{+0.0073}_{-0.0073}$ & $0.37^{+0.26}_{-0.27}$ & $0.0568^{+0.0010}_{-0.0010}$ & $3.27^{+0.27}_{-0.33}$ & $2.64^{+0.84}_{-0.84}$ \\ \vspace{3pt}
TOI-1467.01 & $5.9723^{+0.0012}_{-0.0012}$ & $1766.9828^{+0.0046}_{-0.0046}$ & $0.51^{+0.23}_{-0.27}$ & $0.05077^{+0.00070}_{-0.00070}$ & $1.91^{+0.14}_{-0.14}$ & $1.9^{+1.2}_{-1.2}$ \\ \vspace{3pt}
TOI-1730.01 & $6.226124^{+0.000020}_{-0.000020}$ & $2603.9611^{+0.0014}_{-0.0014}$ & $0.39^{+0.20}_{-0.24}$ & $0.05357^{+0.00069}_{-0.00068}$ & $2.131^{+0.096}_{-0.096}$ & $2.76^{+0.96}_{-0.96}$ \\ \vspace{3pt}
TOI-1730.02 & $2.1555020^{+0.0000062}_{-0.0000064}$ & $2603.3162^{+0.0012}_{-0.0012}$ & $0.34^{+0.21}_{-0.22}$ & $0.02640^{+0.00034}_{-0.00034}$ & $1.442^{+0.077}_{-0.077}$ & $1.48^{+0.11}_{-0.11}$ \\ \vspace{3pt}
TOI-1730.03 & $12.566366^{+0.000038}_{-0.000037}$ & $2598.3690^{+0.0010}_{-0.0011}$ & $0.23^{+0.20}_{-0.19}$ & $0.0855^{+0.0011}_{-0.0011}$ & $2.520^{+0.096}_{-0.102}$ & $2.44^{+0.11}_{-0.11}$ \\ \vspace{3pt}
TOI-1738.01 & $3.69759^{+0.00093}_{-0.00090}$ & $1792.6603^{+0.0053}_{-0.0059}$ & $0.73^{+0.22}_{-0.27}$ & $0.04035^{+0.00042}_{-0.00042}$ & $3.84^{+0.36}_{-0.82}$ & $4.38^{+0.40}_{-0.40}$ \\ \vspace{3pt}
TOI-2120.01 & $5.799792^{+0.000080}_{-0.000080}$ & $1795.8232^{+0.0020}_{-0.0020}$ & $0.79^{+0.16}_{-0.18}$ & $0.0376^{+0.0012}_{-0.0012}$ & $2.52^{+0.26}_{-0.39}$ & $2.63^{+0.39}_{-0.39}$ \\ \vspace{3pt}
TOI-2290.01 & $0.3862296^{+0.0000051}_{-0.0000054}$ & $1764.9858^{+0.0020}_{-0.0020}$ & $0.44^{+0.25}_{-0.27}$ & $0.00855^{+0.00010}_{-0.00010}$ & $1.308^{+0.091}_{-0.091}$ & $1.2^{+2.0}_{-2.0}$ \\ \vspace{3pt}
TOI-2331.01 & $4.715073^{+0.000027}_{-0.000027}$ & $2167.3349^{+0.0033}_{-0.0033}$ & $0.26^{+0.22}_{-0.21}$ & $0.04538^{+0.00055}_{-0.00055}$ & $4.70^{+0.21}_{-0.22}$ & $7.5^{+3.5}_{-3.5}$ \\ \vspace{3pt}
TOI-5561.01 & $2.00760^{+0.00011}_{-0.00011}$ & $2502.1442^{+0.0027}_{-0.0027}$ & $0.79^{+0.22}_{-0.20}$ & $0.02451^{+0.00037}_{-0.00036}$ & $8.3^{+3.2}_{-2.6}$ & $7.2^{+1.5}_{-1.5}$ \\
\enddata
\tablecomments{Updated values are medians and $68\%$ confidence intervals of the posterior distributions. The right-most column contains the previously reported estimate for $R_{\rm p}$. The estimate for EPIC 211510580.01 is from \citet{Kruse2019}. The estimate for EPIC 248775938.01 is from \citet{Castro2020}. The estimates for all of the TOIs are from ExoFOP.}
\tablenotetext{a}{KBJD ({\it Kepler} BJD) applies to {\it K2} targets and is equal to BJD - 2454833. TBJD ({\it TESS} BJD) applies to {\it TESS} targets and is equal to BJD - 2457000.}
\end{deluxetable*}

\begin{figure*}[t!]
    \centering
    \includegraphics[width=0.49\textwidth]{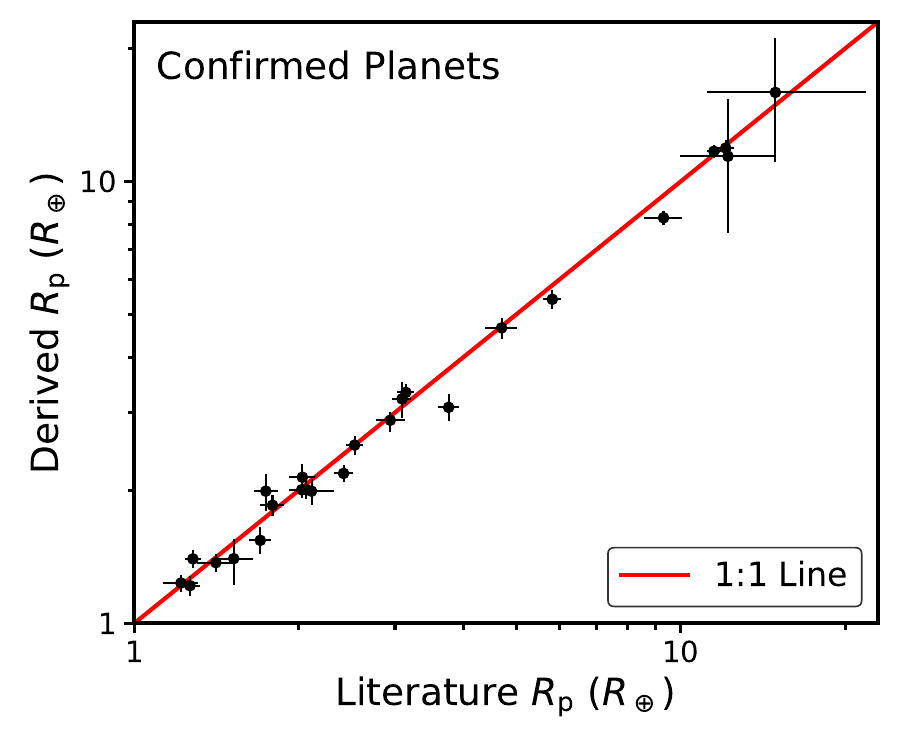}
    \includegraphics[width=0.49\textwidth]{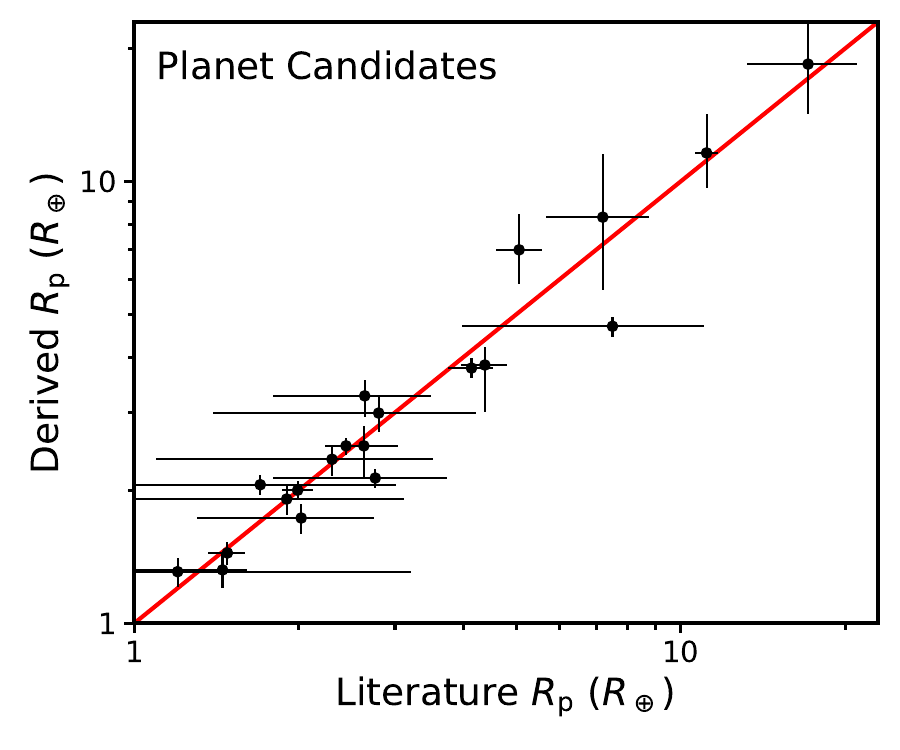}
    \caption{Comparison of planet radii derived in this paper and those previously reported for the confirmed planets in Table \ref{tab: planet_params} (left) and the planet candidates in Table \ref{tab: planet_candidate_params} (right).}
    \label{fig: Rp_comparison}
\end{figure*}

\section{Discussion} \label{sec:discussion}

\subsection{Updated Planet Radii}

A comparison between $R_\mathrm{p}$ values derived from our light curve fits and those listed in the literature are shown in Figure \ref{fig: Rp_comparison}. In general, the values derived here agree with those previously found within $1 \sigma$ uncertainties. However, for a few planets, we derived radii that are discrepant with those listed in the literature at the $1 \sigma$ level, which can significantly influence our understanding of their bulk densities once their masses are measured. 

For one system -- TOI-1201 (TIC 29960110) -- this can be largely attributed to differences in estimates for $R_\star$. For this star, we calculated $R_\star = 0.481 \pm 0.014 \, R_\odot$, whereas \citet{Kossakowski2021} estimates $R_\star = 0.508 \pm 0.016 \, R_\odot$. In their calculation, they determined photospheric properties for the host star following \citet{passegger2019}, which involved comparing a visible wavelength spectrum to a synthetic spectrum, and then determined $R_\star$ using the empirical relations in \citet{schweitzer2019}. This difference in techniques is partially responsible for the $\sim 10\%$ difference in measured planet radii. 

For three other slightly discrepant systems -- TOI-530 (TIC 387690507), TOI-620 (TIC 296739893), and TOI-1468 (TIC 243185500) -- we derived stellar radii consistent with those listed in the planet discovery papers at the $1\sigma$ level. We attribute the differences in measured planet radii to the fact that more \textit{TESS} data became available between the publication of those systems and this work, leading to a transit signal that has a higher S/N and is more robust against systematics. Specifically, \citet{gan2022} did not have access to \textit{TESS} sectors 44 and 45 for their analysis of TOI-530 b, \citet{Reefe2022} did not have access to \textit{TESS} sector 62 for their analysis, and \citep{Charturvedi2022} did not have access to \textit{TESS} sector 57 for their analysis.

\subsection{Stellar Metallicity and System Properties}

To assess the reliability of our results and explore trends between derived stellar parameters and system properties, we compared our results to previously characterized low-mass stars from the NASA Exoplanet Archive (NEA) in Figures \ref{fig: FeH_Ms_Rp} and \ref{fig: FeH_Ms_Np}. We note that only some of the stars from the NEA have properties derived from near-infrared spectra \citep[e.g.,][]{dressing2019}, whereas others have properties derived from photometric relations \citep[e.g.,][]{bonfils2005, Mann2013, kesseli2019} or isochrone fitting \citep[e.g.,][]{huber2014, mathur2017}. Because these properties were not derived in a homogeneous way, there may be systematics present in the data that we do not account for.

Figure \ref{fig: FeH_Ms_Rp} separates stars based on planet size. For stars with planets smaller than $4 \, R_\oplus$, we see that small planets are found around low-mass stars with a range of masses and metallicities in both the NEA sample and our sample. One notable feature is that planet-hosting stars seem to be able to have lower [Fe/H] when $M_\star$ is relatively high, possibly hinting at a poor planet formation efficiency when both [Fe/H] and $M_\star$ are low (Boley et al. submitted). Of the stars in our sample with $M_\star < 0.3 \, M_\odot$, the lowest metallicity star hosting a small planet is TIC 212957629 (TOI-2406), with ${\rm [Fe/H]} = -0.492 \pm 0.088$ dex. Of the more massive planet-hosting stars in our sample, TIC 170849515 (TOI-555) and TIC 163539739 (TOI-1278) reach considerably lower metallicities of ${\rm [Fe/H]} = −0.169 \pm 0.137$ dex and ${\rm [Fe/H]} = −0.182 \pm 0.086$ dex, respectively. However, we cannot rule out a sample selection bias as the cause of this apparent trend.

\begin{figure*}[t!]
    \centering
    \includegraphics[width=\textwidth]{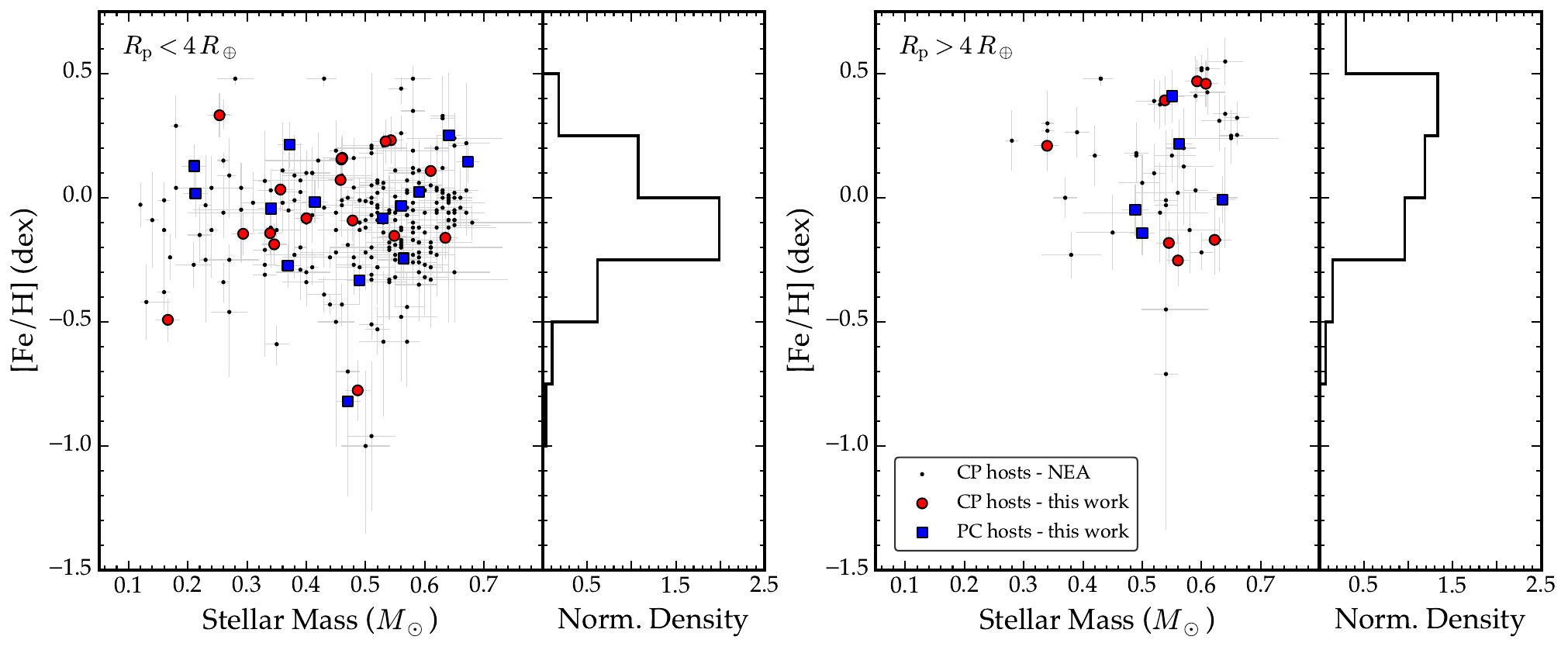}
    \caption{Distribution of stellar masses and metallicities for low-mass stars with at least one small planet ($R_{\rm p} < 4 \, R_\oplus$; left) or at least one giant planet ($R_{\rm p} > 4 \, R_\oplus$; right). Black dots are the masses and metallicities of stars with $T_{\rm eff} < 4300$ K, $\log g > 4$, and $M_\star < 0.7 \, M_\odot$ collected from the NASA Exoplanet Archive, which were calculated using a variety of techniques. Red circles are stars in our sample with at least one confirmed planet. Blue squares are stars in our sample with no confirmed planets but at least one planet candidate. The grey bars are $1\sigma$ uncertainties. The histograms on the right of each panel are the normalized density distributions of all points combined. In general, our measured values agree well with previously reported values of similar systems.}
    \label{fig: FeH_Ms_Rp}
\end{figure*}

\begin{figure*}[t!]
    \centering
    \includegraphics[width=\textwidth]{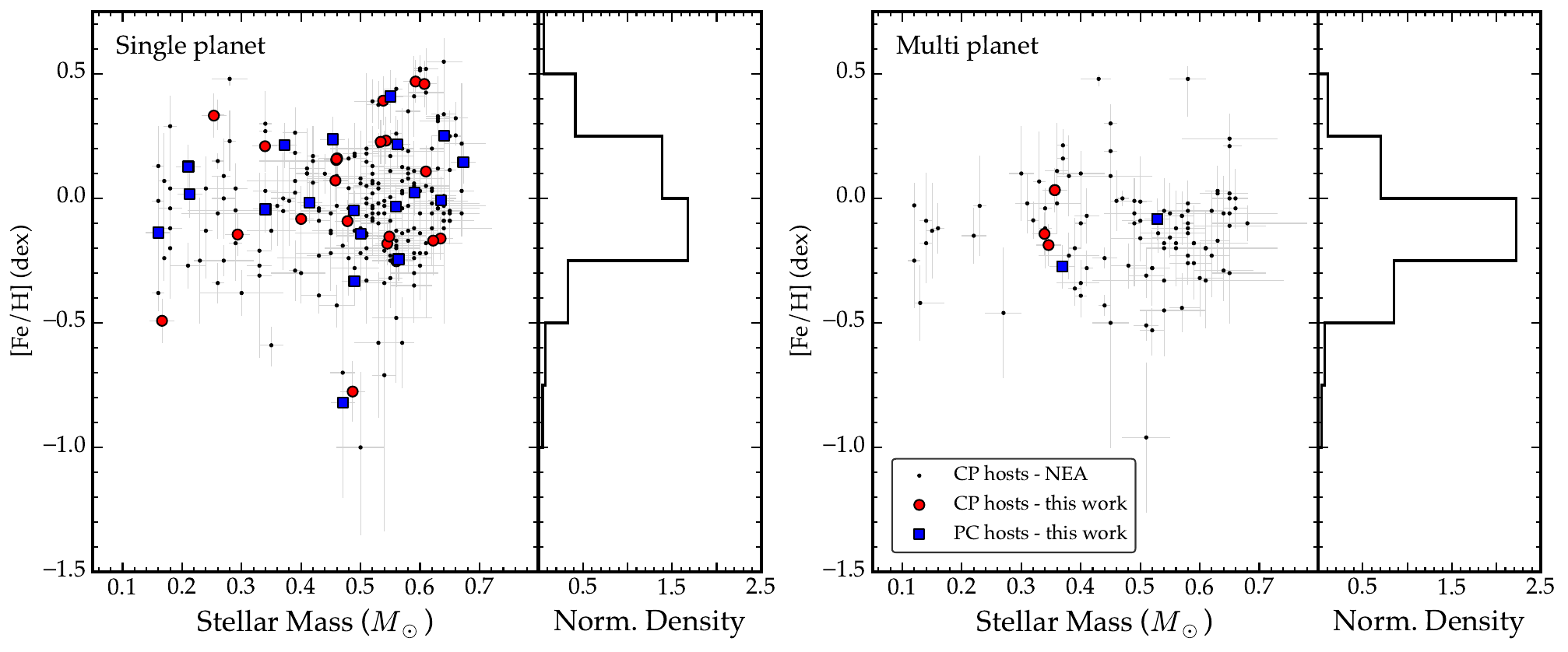}
    \caption{The same as Figure \ref{fig: FeH_Ms_Rp}, but for systems with only one detected planet or planet candidate (left) and systems with multiple detected planets or planet candidates (right).}
    \label{fig: FeH_Ms_Np}
\end{figure*}

Compared to those with small planets, low-mass stars with planets larger than $4 \, R_\oplus$ appear to occupy a distinctly high-[Fe/H] and high-$M_\star$ region of parameter space in both the NEA sample and our sample. However, determining if this tendency is physical would require a more rigorous analysis than we attempt here. Hot Jupiters are known to be rare around stars of nearly all spectral types \citep[e.g.,][]{zhou2019, beleznay2022, gan2023occurrence, zink2023}, and late M dwarfs are notoriously difficult to search for and confirm planets around due to their low brightnesses at visible wavelengths. Determining if hot Jupiters are truly absent around stars with the lowest masses would require one to assess the search completeness of these stars by \textit{Kepler}, \textit{K2}, and \textit{TESS}. The tendency for giant planets to orbit more metal-rich stars is well known for FGK-type stars \citep[e.g.,][]{fischer2005, johnson2010, sousa2011}, but as of yet this trend has not been identified for M dwarfs with a high statistical significance \citep[e.g.,][]{gaidos2014}. Identifying this trend confidently will likely require a large sample of low-mass planet-hosting stars with uniformly derived metallicities.

Figure \ref{fig: FeH_Ms_Np} contains the same stellar sample, but separates the stars based on planet multiplicity. The [Fe/H] distributions of the two panels are qualitatively similar, although a slight bias towards lower metallicities appears to be present for the stars with multiple known planets, consistent with the findings of \citet{anderson2021} and \citet{rodriguezmartinez2023}. However, we again note that this figure contains stars with metalliticites derived using a variety of methods, so this apparent tendency may not be significant. We also note that many of the single planet systems may have additional close-in planets that have yet to be detected, either due to a limited photometric baseline or because the additional planets are non-transiting and the stars have not been monitored with radial velocities. As \textit{TESS} continues to collect data, and low-mass stars with transiting planets are targeted by radial velocity surveys, the relationship between stellar metallicity and planet multiplicity will likely become clearer.

\section{Conclusions} \label{sec:conclusions}

We present the results of a survey to collect near-infrared spectra of cool planet-hosting stars with SpeX on the IRTF. We collected spectra of 65 cool stars in order to measure their metallicities and calculate updated stellar parameters. Twenty five of these stars have confirmed transiting planets and 27 have transiting planet candidates. Five of our targets host multiple confirmed planets or multiple planet candidates. The remaining stars either have no planet candidates or previously had planet candidates that have since been determined to be false positives.

Using our metallicities and updated stellar parameters, we performed updated fits of the \textit{K2} and \textit{TESS} light curves to calculate updated properties and orbital parameters of the confirmed planets and planet candidates, finding results in general agreement with those reported earlier. We also examine the distribution of planets of different sizes as a function of stellar properties. We found that our derived stellar parameters are consistent with previously studied stars that host similar planetary systems.

Our sample increases the number of planet-hosting M dwarfs and late K dwarfs with spectroscopically determined metallicities. As more planets around cool stars are discovered, these metallicities will enable more rigorous investigations of the correlations between planetary systems and the properties of their low-mass stars.

\acknowledgements

The authors thank Elisabeth R. Newton and Andrew Vanderburg for their constructive feedback on this work and the telescope proposals associated with the data presented herein.

SG is supported by an NSF Astronomy and Astrophysics Postdoctoral Fellowship under award AST-2303922. SG and CDD acknowledge support from NASA FINESST award 80NSSC20K1549. CDD and ET also acknowledge support provided by the David and Lucile Packard Foundation via grant 2019-69648, the NASA TESS Guest Investigator Program via grant 80NSSC18K1583, and the NASA Exoplanets Research Program (XRP) via grant 80NSSC20K0250. JZ acknowledges support provided by NASA through the Hubble Fellowship grant HST-HF2-51497.001 awarded by the Space Telescope Science Institute, which is operated by the Association of Universities for Research in Astronomy, Inc., for NASA, under the contract NAS 5-26555.

This paper includes data collected at the Infrared Telescope Facility, which is operated by the University of Hawaii under contract 80HQTR19D0030 with the National Aeronautics and Space Administration. This research has made use of the NASA Exoplanet Archive, which is operated by the California Institute of Technology, under contract with the National Aeronautics and Space Administration under the Exoplanet Exploration Program. This research has made use of the Exoplanet Follow-up Observation Program (ExoFOP; DOI: 10.26134/ExoFOP5) website, which is operated by the California Institute of Technology, under contract with the National Aeronautics and Space Administration under the Exoplanet Exploration Program.

\software{\texttt{exoplanet} \citep{exoplanet2021}, \texttt{Lightkurve} \citep{lightkurve2018}, \texttt{metal} \citep{Mann2013}, \texttt{pymc3} \citep{Salvatier2016}, \texttt{Spextool} \citep{2004PASP..116..362C}, \texttt{tellrv} \citep{Newton2014,Newton2022}, \texttt{w{\={o}}tan} \citep{hippke2019}, \texttt{xtellcor} \citep{2003PASP..115..389V}}

\facilities{IRTF, Exoplanet Archive, ExoFOP, {MAST (HLSP)}}

\bibliography{bibliography}{}
\bibliographystyle{aasjournal}

\end{document}